\documentclass[11pt]{article}
\usepackage[utf8]{inputenc}
\usepackage{amsmath}
\usepackage{amsfonts}
\usepackage{amssymb}
\usepackage{graphicx}
\usepackage{siunitx}
\usepackage{xcolor}
\usepackage{cite}
\usepackage{tikz}
\usepackage{bm,dsfont} 
\usepackage{float}
\usepackage{pgfplots}
\usepgfplotslibrary{fillbetween}
\usepackage{url}
\usepackage{siunitx}
\usepackage{hyperref}
\usepackage{xcolor}
\usepackage{array,booktabs}
\hypersetup{linktocpage,pdfauthor={Omar},colorlinks=true,urlcolor=purple,linkcolor=purple,citecolor=purple}
\pgfplotsset{ compat = newest}
\usepgfplotslibrary{external}
\newcommand{\lambdabar}{{\mkern0.75mu\mathchar '26\mkern -9.75mu\lambda}}
\usetikzlibrary{
	babel,
	quotes,
	arrows.meta,
	shapes.symbols,
	shadows.blur,
}

\tikzset{
	> = LaTeX,
	pics/lisa/.style = {
		code = {
			\draw [densely dashed, thick] (0, 1) -- (210:1) -- (330:1) -- cycle;
			\foreach \th in {90, 210, 330} {
				\draw [thick, fill = white] (\th:1) circle [radius = 0.2];
			}
		}
	}
}

\definecolor{lime}{HTML}{A6CE39}
\DeclareRobustCommand{\orcidicon}{
	\begin{tikzpicture}
	\draw[lime, fill=lime] (0,0) 
	circle [radius=0.16] 
	node[white] {{\fontfamily{qag}\selectfont \tiny ID}};
	\draw[white, fill=white] (-0.0625,0.095) 
	circle [radius=0.007];
	\end{tikzpicture}
	\hspace{-2mm}
}

\foreach \x in {A, ..., Z}{%
	\expandafter\xdef\csname orcid\x\endcsname{\noexpand\href{https://orcid.org/\csname orcidauthor\x\endcsname}{\noexpand\orcidicon}}
}

\numberwithin{equation}{section}

\begin{document}
	
\title{ \textbf{Gravitational Waves in Einstein-Cartan Theory:\\ On the Effects of Dark Matter Spin Tensor}}
\author{Emilio Elizalde$^{1}$\orcidE{},
Fernando Izaurieta$^{2}$\orcidF{},
Cristian Riveros$^{3}$,
Gonzalo Salgado$^{4}$ \\ and Omar Valdivia$^{5,6}$\orcidA{}.
\bigskip \\
{\small \textit{$^{1}$ Institute of Space Sciences (IEEC-CSIC), Campus Universitat Autònoma de Barcelona,}} \\{ \small  \textit{ C. Can Magrans s/n, 08193 Bellaterra (Barcelona), Spain.}}\\
{\small \textit{$^{2}$ Facultad de Ingenier\'{\i}a, Arquitectura y Dise\~{n}o, Universidad San Sebasti\'{a}n,}} \\{ \small  \textit{ Lientur 1457, Concepci\'{o}n 4080871, Chile.}}\\
{\small \textit{$^{3}$ Departamento de F\'{\i}sica, Universidad Nacional de Colombia, 111321 Bogot\'{a}, Colombia.}}\\
{\small \textit{$^{4}$ Departamento de F\'{\i}sica, Universidad de Concepci\'{o}n, casilla 160-C, Concepci\'{o}n, Chile.}}\\
{\small \textit{$^{5}$ Instituto de Ciencias Exactas y Naturales ICEN, Universidad Arturo Prat, Iquique, Chile.}}\\
{\small \textit{$^{6}$ Facultad de Ciencias FdeC, Universidad Arturo Prat, Iquique, Chile.}}}
\maketitle

\begin{abstract}
This article studies the effects of an arbitrary dark matter spin tensor on the propagation of gravitational wave amplitude in the context of Einstein-Cartan theory.
We choose to work with an arbitrary spin tensor because, given our ignorance of the nature of dark matter, it is sensible not to make further hypotheses on its spin and not to assume any particular model for its spin tensor (or its vanishing).
The analysis focuses on a \textquotedblleft weak-torsion regime,\textquotedblright\ such that gravitational wave emission, at leading and subleading orders, does not deviate from standard General Relativity. We show that, in principle, the background torsion induced by an eventual dark matter spin component could lead to an anomalous dampening or amplification of the gravitational wave amplitude, after going across a long cosmological distance.
We assess the importance of this torsion-induced anomalous amplitude propagation for binary black hole mergers in a way as model-free as possible in Einstein-Cartan gravity. It is possible to prove that at its best, for realistic late-universe cosmological scenarios, the effect is tiny and falls below detection thresholds, even for near-future interferometers such as LISA. Therefore, detecting this effect may not be impossible, but it is still beyond our technological capabilities. As a model-independent result in the Einstein-Cartan context, it also implies that mergers are robust standard sirens without considering any potential dark-matter-induced torsional effects.

\end{abstract}
\newpage
\tableofcontents
\newpage
\section{Introduction}

This article will assess whether unaccounted torsion effects in Einstein-Cartan-Sciama-Kibble (ECSK) theory of gravity, could introduce systematic errors when using binary black hole mergers as standard sirens and whether mergers could help falsify the presence of ECSK torsion components.

This problem is crucial for understanding the universe's evolution and exploring the nature of dark energy and dark matter. Both, dark energy and dark matter, give rise to significant physical effects, such as the accelerated expansion of the Universe ~\cite{SupernovaSearchTeam:1998fmf,SupernovaCosmologyProject:1998vns,SupernovaSearchTeam:2003cyd,SupernovaCosmologyProject:2003dcn}, gravitational lensing around galaxy clusters~\cite{Navarro:1995iw}, and the anomalous velocity profile of stars orbiting spiral galaxies~\cite{1978ApJ225L.107R}, among several other effects.

The best observational evidence for dark energy is the so-called distance-redshift relation, coming from the observation of Type Ia supernovae (SN). They seem robust standard candles, and it is possible to calibrate their observed brightness and luminosity distance. However, the absence of a solid theoretical description still leaves open the possibility of having, for instance, evolutionary processes in SN brightness leading to unknown systematic errors and consequently threatening the confidence in the estimation of cosmological parameters~\cite{Drell_2000}.

In contrast, gravitational waves (GW) and multi-messenger astronomy promise an era of high-precision cosmology~\cite{Abbott_2017,Baker:2017hug,Visinelli:2017bny,Casalino:2018tcd,Odintsov:2020sqy,Shiralilou:2021mfl,East:2020hgw,Bernard:2022noq,Oikonomou:2022ksx,Mezzasoma:2022pjb,Liu:2022qcx,Ezquiaga:2017ekz,Sakstein:2017xjx,Creminelli:2017sry}. In particular, the GW-driven spiraling dynamics of binary black holes (BBH) may provide a way for high-accuracy measurements of luminosity distances, $\delta D_{L}/D_{L}\sim1-10\%$. Moreover, in a multi-messenger event, the electromagnetic counterpart measurement of the redshift~\cite{1986Natur.323..310S} could allow us to reduce this error even further to $0.5-1\%$. For this reason, BHH mergers as standard sirens are essential for developing high-precision cosmology~\cite{Holz:2005df}.

All this is particularly true for high redshifts. BBH merger events should follow the mergers of galaxies and pregalactic structures at high redshift~\cite{2003ApJ...582..559V}. Though the merger rate is poorly understood, LISA should measure at least several events over its mission, especially considering its sensitivity~\cite{2017arXiv170200786A}. For this reason, it is crucial to examine any source of systematic error that could arise when using BBH GW as standard sirens.

This article assesses whether a possible source of systematic error could arise from not considering torsion when studying the propagation of GW. Torsion is not such a far-fetched possibility; for instance, dark matter could be a source of torsion, and torsion could be a dark matter component~\cite{Tilquin:2011bu,Alexander:2019wne,Magueijo:2019vmk,Barker:2020gcp,Alexander:2020umk,Izaurieta:2020xpk}.

Of course, there are multiple proposals for dark matter candidates, such as weakly interacting massive particles, sterile neutrinos, axions, cold massive halo objects, and primordial black holes~\cite{Gelmini:2015zpa,Bertone:2004pz,Young2016}, among others. Moreover, due to the standard cosmology hassles for satisfactorily describing dark matter (see, for instance,~\cite{Bull:2015stt}), there is an active trend for studying modified gravity theories~\cite{Elizalde:2004mq,Briscese:2006xu,Cognola:2006eg,Cognola:2007zu,Cognola:2008zp,Elizalde:2009gx,DeFelice:2010aj,Elizalde:2010jx,Olmo:2011uz,Capozziello:2011et,Nojiri:2017ncd,BeltranJimenez:2017doy,Heisenberg:2018vsk,Corman:2022xqg,Sennett:2016klh}.

In general, many of these dark matter candidates can give rise to torsion. For instance, when considering modified gravity theories, non-minimal couplings and second derivatives in the Lagrangian are actual sources of torsion~\cite{Barrientos:2019awg}. Furthermore, when considering the dark matter as new particles, it is important to stress that fermions give rise to torsion. In particular, this article will focus on particle dark matter giving rise to torsion, and the consequences of this torsion on GW propagation (and BBH as standard sirens).

To analyze dark matter particles as a possible source of torsion, we use ECSK gravity~\cite{Trautman:2006fp}. In ECSK, quantum mechanical spin acts as the source of torsion\footnote{It is important to stress that with spin, we refer only to intrinsic quantum mechanical spin, and we should not confuse this with the angular momentum density. This natural confusion has already led to some mistakes in the literature, see Ref.~\cite{Hehl:2013qga}}, in the same way as energy is a source of curvature~\cite{Blagojevic:2013xpa,Arcos:2005ec}.  Consequently, torsion could have been relevant in the early universe because of its extremely high fermion densities~\cite{Poplawski:2011jz,Unger:2018oqo,Kranas:2018jdc,Poplawski:2010kb,Ivanov:2016xjm,Razina:2010bj,Palle:2014goa,Poplawski:2012qy,Cubero:2019lxw, Poplawski:2020hrp, Poplawski:2020kzc}. Moreover, in standard ECSK the torsion does not propagate in a vacuum, and it interacts very weakly with Standard Model fermions (see Chap. 8.4 of Ref.~\cite{SupergravityVanProeyen} and Ref.~\cite{Puetzfeld:2014sja,Carroll:1994dq,Boos:2016cey}). Thus, torsion might potentially be a component of dark matter~\cite{Tilquin:2011bu,Alexander:2019wne,Magueijo:2019vmk,Barker:2020gcp,Alexander:2020umk,Izaurieta:2020xpk}.

However, analyzing dark matter as a potential torsion source is difficult. Given our ignorance of the fundamental nature of dark matter, we do not have any clue about its spin. Moreover, we do not know whether dark matter's spin tensor has non-zero components or if they identically vanish. Without the spin tensor phenomenology, we are clueless about the value of its associated torsion.

In the current article, we will deal with this difficulty and perform an analysis as model-independent as possible. To keep the conclusions as general as possible, on purpose, we will not use any dark matter spin tensor model. For instance, current ECSK literature has thoroughly examined standard model spin-1/2 fermions following the Dirac equation as sources of torsion. We want to remark that we will not use this as a dark matter model because it is a non-starter. On the one hand, it seems extremely unlikely that any standard model particle could explain dark matter phenomenology. On the other hand, we already know that this kind of particle requires enormous spin densities to produce a significant torsion (See Ref.~\cite{Puetzfeld:2014sja,Carroll:1994dq,Boos:2016cey}). Therefore, in any astrophysically realistic phenomenon in the late universe, the spin tensor produced by standard model spin-1/2 fermions is irrelevant and will not affect GW propagation at leading and subleading order.

Given the correct mathematical tools (see Sec.~\ref{SecII}), it is possible to reach significant results without invoking particular spin tensor models. For instance, it is possible to prove that for any theory based on Riemann-Cartan geometry, the Einstein-Hilbert term does not change the dispersion relation, i.e., the speed of GW does not change.  However, torsion may influence the propagation of the GW amplitude and polarization (see Ref.~\cite{Barrientos:2019awg}). In principle, these effects impact directly the reliability of mergers as standard sirens. 

However, this work proves, without invoking any particular spin tensor model, that the anomalous amplitude propagation effect (due to ECSK torsion compatible with cosmological observations) falls several orders of magnitude below the observation threshold of future LISA-like observatories. Therefore, we conclude that in practice, BH mergers are not good enough to falsify ECSK torsion components, and they are reliable standard sirens even without considering any potential ECSK torsion effects. Furthermore, these results are very robust: we arrive at them without introducing any particular model for the spin tensor when $z<1$ and therefore are model-independent in the ECSK context. Consequently, we can expect that results of standard torsionless GR will still be valid regarding amplitude propagation, regardless of whether the dark matter has a nonvanishing spin tensor.
	
This spin-tensor-model-independent proof uses one central hypothesis to make its predictions. To agree with GW data, we assume a weak torsional background scenario. This means that GW emission occurs as in GR; with torsional effects being negligible at the subleading order in the eikonal limit at the moment of it emission.  
 To make estimations at $z>1$, we also use some popular cosmological ECSK torsional models that are compatible with observations. However, results remain the same: they all predict minor amplitude anomalies that fall several orders of magnitude below the observation threshold of future LISA-like observatories.

This article has the following structure: In section \ref{SecII}, we briefly review the essential aspects of i) how to properly implement wave operators over Riemann-Cartan geometries, ii) linear and second-order perturbations of fields and their potential contributions at leading and sub-leading order in the eikonal approximation\footnote{There have been other approaches for studying GW beyond Riemannian geometry (See for instance~\cite{Obukhov:2017pxa, Jimenez-Cano:2022arz} and references therein). The advantage of our analysis relies on the fact that a proper definition of wave operators on Riemann-Cartan spaces enables us to carry out the wave propagation analysis to every order in the eikonal approximation instead of limiting ourselves to the leading one.}. Then, in section \ref{SecIII}, we study GW propagation in ECSK theory, and we explicitly compute the anomalous propagation of the amplitude and polarization of GW in the eikonal approximation.
In section {\ref{SecIV}}, we discuss the possibility of using GW for studying the implications of torsion at cosmological scales, and we constrain different ansatz used in literature based on observational data. Finally, section \ref{SecV} contains the summary and conclusions of the paper. 
	
\section{Waves in Riemann-Cartan geometry}
\label{SecII}
\subsection{Notation}

When describing a theory on a Riemann-Cartan geometry (i.e., without imposing
the vanishing torsion condition), there are several alternatives regarding
mathematical language and notation. The two most common alternatives are (1)
differential forms on an orthonormal basis, i.e., describing geometry in terms
of the vielbein 1-form $e^{a}=e^{a}{}_{\mu}\mathrm{d}x^{\mu}$ and the spin
connection 1-form $\omega^{ab}=\omega^{ab}{}_{\mu}\mathrm{d}x^{\mu}$, and (2) the
standard tensorial language, describing the geometry in terms of the metric
$g_{\mu \nu}$ and the affine connection $\Gamma_{\mu \nu}^{\lambda}$ on the
coordinate basis.

Of course, the choice of language and basis is physically and mathematically
irrelevant. Nevertheless, a particular basis can be more
advantageous/traditional to express some ideas than others. For this reason, we
will use the conciseness of the differential form language when referring to the
general mathematical properties of the wave operator on a Riemann-Cartan
geometry (Sec.~\ref{SecII}). However, we will use traditional tensor components in
the standard coordinate basis when analyzing gravitational waves'
propagation on a cosmological background (Sec.~\ref{SecIII}).

We consider a $4$-dimensional spacetime manifold $M$ with $\left(
-,+,+,+\right)  $ signature. Let us use lowercase Greek characters to denote
elements of the coordinate basis of vectors $\left \{  \partial_{\mu}\right \}
$ and 1-forms $\left \{  \mathrm{d}x^{\mu}\right \}  $, and lowercase Latin
characters to denote elements of the orthonormal basis of vectors $\left \{
\boldsymbol{e}_{a}=e_{a}{}^{\mu}\partial_{\mu}\right \}  $ and 1-forms
$\left \{  e^{a}=e^{a}{}_{\mu}\mathrm{d}x^{\mu}\right \}  $. 

The Lorentz curvature and torsion 2-forms are defined by%
\begin{equation}
R^{ab}=\mathrm{d}\omega^{ab}+\omega^{a}{}_{c}\wedge \omega^{cb}\,,
\end{equation}
\begin{equation}
T^{a}=\mathrm{D}e^{a}=\mathrm{d}e^{a}+\omega^{a}{}_{b}\wedge e^{b}\,,
\end{equation}
where $\mathrm{d}:\Omega^{p}\left(  M\right)  \rightarrow \Omega^{p+1}\left(
M\right)  $ denotes the exterior derivative and $\mathrm{D}=\mathrm{d}+\omega$\,,
denotes the Lorentz-covariant derivative.

A circle above an entity denotes the torsionless version of it. For instance, it is possible to write the spin connection 1-form as
\begin{equation}
\omega^{ab}=\mathring{\omega}^{ab}+\kappa^{ab}\,,
\end{equation}
such that
$\kappa^{ab}$ is the anti-symmetric one-form contorsion (or contortion),%
\begin{equation}
T^{a}=\kappa^{a}{}_{b}\wedge e^{b}\,.
\end{equation}

\subsection{Lichnerowicz-DeRham wave operator on a Riemann-Cartan geometry}

Ref.~\cite{Barrientos:2019msu} provides a formal mathematical definition of the Lichnerowicz-DeRham wave operator on a Riemann-Cartan geometry, and in Ref.~\cite{Valdivia:2017sat,Barrientos:2019awg} we find examples of using it in theories with nonvanishing torsion. We refer to these former articles for a deeper treatment, but nevertheless this section briefly reviews the wave operators definitions and properties in terms of the differential form language. 

In terms of the $d=4$ Hodge-dual operator $\ast:\Omega^{p}\left(  M\right) \rightarrow \Omega^{4-p}\left(  M\right)  $, we define the operator $\mathrm{I}_{a_{1}\cdots a_{q}}:\Omega^{p}\left(  M\right)  \rightarrow \Omega^{p-q}\left(  M\right)$, on the 4-dimensional spacetime manifold $M$ as
\begin{equation}
\mathrm{I}_{a_{1}\cdots a_{q}}=-\left(  -1\right)  ^{p\left(  p-q\right)}\ast \left(  e_{a_{1}}\wedge \cdots \wedge e_{a_{q}}\wedge \ast \right.\,.
\end{equation}
At this point, it is useful to define two new derivatives on Riemann-Cartan
geometries.
The first of them is the generalized covariant coderivative $\mathrm{D}^{\ddag}:\Omega^{p}\left(  M\right)  \rightarrow \Omega^{p-1}\left(  M\right)$ as
\begin{equation}
\mathrm{D}^{\ddag}=-\mathrm{I}^{a}\mathrm{DI}_{a}\,.
\end{equation}
The second one corresponds to $\mathcal{D}_{a}:\Omega^{p}\left(  M\right) \rightarrow \Omega^{p}\left(  M\right)  $ defined by
\begin{equation}
\mathcal{D}_{a}=\mathrm{I}_{a}\mathrm{D}+\mathrm{DI}_{a}\,.
\end{equation}
The operator $\mathrm{D}^{\ddag}$ is the Riemann-Cartan generalization of the standard coderivative operator $\mathrm{d}^{\dag}=\ast \mathrm{d}\ast$ of Riemannian geometry. When torsion vanishes, $\mathrm{D}^{\ddag}$ and $\mathrm{d}^{\dag}$ coincide.
The derivative $\mathcal{D}_{a}$ satisfies the Leibniz rule, and it generalizes the standard coordinate definition $\nabla=\partial+\Gamma$ in such a way that $-\mathcal{D}_{a}\mathcal{D}^{a}$ defines the generalized \textit{(torsionfull)} Beltrami wave operator. In fact, it is possible to prove that
\begin{equation}
    \mathcal{D}_{a}=e_{a}{}^{\mu}\nabla_{\mu}+\mathrm{I}_{a}T^{b}\wedge \mathrm{I}_{b}\,,\label{Eq_D=D}
\end{equation}
and therefore their vanishing torsion counterparts $\mathcal{\mathring{D}}_{a}$ and $\mathring{\nabla}_{\mu}$ are equivalent.
A practical feature of operators $\mathrm{I}_{a},\mathrm{D,}$ and
$\mathcal{D}_{a}$ is that they obey the Leibniz rule, and they span an open
superalgebra satisfying the super Jacobi identity (see Ref.~\cite{Barrientos:2019msu}):%
\begin{align}
\left \{  \mathrm{I}_{a},\mathrm{D}\right \}   &  =\mathcal{D}_{a}\,, \label{os1}\\
\left \{  \mathrm{I}_{a},\mathrm{I}_{b}\right \}   &  =0\,,\\
\left \{  \mathrm{D},\mathrm{D}\right \}   &  =2\mathrm{D}^{2}\\
\left[  \mathrm{I}_{a},\mathcal{D}_{b}\right]   &  =-T^{c}{}_{ab}%
\mathrm{I}_{c}\,,\\
\left[  \mathrm{D},\mathcal{D}_{a}\right]   &  =\mathrm{D}^{2}\mathrm{I}%
_{a}-\mathrm{I}_{a}\mathrm{D}^{2}\,,\\
\left[  \mathcal{D}_{a},\mathcal{D}_{b}\right]   &  =\mathrm{I}_{ab}%
\mathrm{D}^{2}+\mathrm{D}^{2}\mathrm{I}_{ab}+\mathrm{I}_{a}\mathrm{D}%
^{2}\mathrm{I}_{b}-\mathrm{I}_{b}\mathrm{D}^{2}\mathrm{I}_{a}-\left(
\mathrm{D}T^{c}{}_{ab}\wedge \mathrm{I}_{c}+T^{c}{}_{ab}\mathcal{D}_{c}\right) 
\,. \label{os2}
\end{align}
These properties greatly simplify algebraic calculations.

In terms of $\mathrm{D}^{\ddag}$, the Lichnerowicz-DeRham wave operator
corresponds to $\square=\mathrm{D}^{\ddag}\mathrm{D}+\mathrm{DD}^{\ddag}$ on
Riemann-Cartan geometries. From a mathematical point of view, it is the right
definition because it satisfies a generalized Weitzenb\"{o}ck identity (see Ref.~\cite{Barrientos:2019msu})%
\begin{equation}
\mathrm{D}^{\ddag}\mathrm{D}+\mathrm{DD}^{\ddag}=-\mathcal{D}^{a}%
\mathcal{D}_{a}+\mathrm{I}_{a}\mathrm{D}^{2}\mathrm{I}^{a}%
\,,\label{Eq_Weitzenboeck}%
\end{equation}
where the second term gives rise to the Lorentz curvature via the Bianchi
identities. From a physical and phenomenological point of view,
Eq.~(\ref{Eq_Weitzenboeck}) is also the appropriate wave operator definition.
It is because $\mathrm{D}^{\ddag}\mathrm{D}+\mathrm{DD}^{\ddag}$ (or
equivalently, the generalized Beltrami operator $-\mathcal{D}^{a}%
\mathcal{D}_{a}$) is the wave operator that arises from perturbations of the
Einstein-Hilbert term in the case of nonvanishing torsion.

\subsection{Generic perturbations of Riemann-Cartan geometry}

The Ref.~\cite{Izaurieta:2019dix} considered general perturbations of a Riemann-Cartan geometry.
Here we briefly review these results to apply them to gravitational waves in the
Einstein-Cartan theory.

In a Riemann-Cartan geometry, the vierbein and the spin connection correspond
to independent degrees of freedom, and in consequence, their perturbations
are, too. It proves convenient to write these independent perturbations as
\begin{align}
e^{a}\mapsto \bar{e}^{a} &  =e^{a}+\frac{1}{2}H^{a},\label{Ec_e-H}\\
\omega^{ab}\mapsto \bar{\omega}^{ab} &  =\omega^{ab}+U^{ab}\left(  H\right)
+V^{ab}\,.\label{Ec_omega_U_V}%
\end{align}
Here, the $H^{a}=H^{a}{}_{b}e^{b}$ 1-form maps the same degrees of freedom as
the standard metric perturbation $h_{\mu \nu}$. The term $U^{ab}\left(  H\right)  $ describes the connection piece that depends on $H^{a}$ and its derivatives through
\begin{equation}
U^{ab}\left(  H\right)  =U_{\left(  1\right)  }^{ab}+U_{\left(  2\right)
}^{ab}+U_{\left(  3\right)  }^{ab}+\cdots
\end{equation}
where%
\begin{align}
U_{\left(  1\right)  }^{ab} &  =-\frac{1}{2}\left(  \mathrm{I}^{a}%
\mathrm{D}H^{b}-\mathrm{I}^{b}\mathrm{D}H^{a}\right)  \,,\\
U_{\left(  2\right)  }^{ab} &  =\frac{1}{8}\mathrm{I}^{ab}\left(
\mathrm{D}H_{c}\wedge H^{c}\right)  -\frac{1}{2}\left[  \mathrm{I}^{a}\left(
U_{\left(  1\right)  }^{bc}\wedge H_{c}\right)  -\mathrm{I}_{b}\left(
U_{\left(  1\right)  }^{ac}\wedge H_{c}\right)  \right]  \,,
\end{align}
and (1), (2) label linear and quadratic terms in $H^{a}$. For a theory
with propagating torsion, the contorsional perturbation term $V^{ab}=V^{ab}%
{}_{c}e^{c}$ 1-form is an independent degree of freedom, and it describes a
`roton' or `torsionon' (see Ref.~\cite{Izaurieta:2019dix,Hehl1980,Boos:2016cey}). For non-propagating torsion theories (as
Einstein-Cartan theory), it is possible to solve $V^{ab}$ in terms of $H^{a}$
and $T^{a}$, in such a way that in a region where the background torsion
vanishes, the perturbation $V^{ab}$ also does.
Up to second order, the perturbations of Lorentz curvature and torsion read%
\begin{align}
T_{a} &  \mapsto \bar{T}_{a}=T_{a}+T_{a}^{\left(  1\right)  }+T_{a}^{\left(
	2\right)  }\,,\\
R^{ab} &  \mapsto \bar{R}^{ab}=R^{ab}+R_{\left(  1\right)  }^{ab}+R_{\left(
	2\right)  }^{ab}\,,
\end{align}
where%
\begin{align}
T_{\left(  1\right)  }^{a} &  =V^{a}{}_{b}\wedge e^{b}-\frac{1}{2}%
\mathrm{I}^{a}\left(  H^{b}\wedge T_{b}\right)  \,,\label{Eq_T(1)}\\
T_{\left(  2\right)  }^{a} &  =\frac{1}{2}V^{a}{}_{b}\wedge H^{b}+\frac{1}%
{4}\mathrm{I}^{a}\left[  H^{b}\wedge \mathrm{I}_{b}\left(  H^{c}\wedge
T_{c}\right)  \right]  \,,\label{Eq_T(2)}\\
R_{\left(  1\right)  }^{ab} &  =\mathrm{D}U_{\left(  1\right)  }%
^{ab}+\mathrm{D}V^{ab}\,,\label{Eq_R(1)}\\
R_{\left(  2\right)  }^{ab} &  =\mathrm{D}U_{\left(  2\right)  }^{ab}+\left(
U_{\left(  1\right)  c}^{a}+V^{a}{}_{c}\right)  \wedge \left(  U_{\left(
	1\right)  }^{cb}+V^{cb}\right)  \,.\label{Eq_R(2)}%
\end{align}

In order to consider these perturbations a proper gravitational wave, they must satisfy
the eikonal condition, i.e., they must correspond to variations in a
characteristic scale $\lambda$ much shorter than the characteristic scale $L$
of changes on the background geometry. This condition defines the eikonal parameter 
\begin{equation}
\epsilon=\frac{\lambdabar}{L}\ll1\,. \label{Eq_Eikonal_Parameter}
\end{equation}
Furthermore, it is possible to write the perturbations $H^{a}$ and
$V^{ab}$ as%
\begin{align}
H^{a} &  =\mathrm{e}^{i\theta}\sum \limits_{p=0}^{\infty}\mathbb{H}_{\left(p\right)  }^{a}\,,\label{Eq_H_eikonal_expan}\\
V^{a}{}_{b} &  =\mathrm{e}^{i\theta}\sum \limits_{p=0}^{\infty}\mathbb{V}_{\left(  p\right)  }^{ab}\,,\label{Eq_V_eikonal_expan}
\end{align}
where $\theta$ corresponds to a real, rapidly-changing phase, and its
derivative defines the wavefront 1-form $k=k_{\mu}\mathrm{d}x^{\mu}%
=\mathrm{d}\theta\sim 1/\lambdabar$ as the dual of the wave vector. In the summation, the terms
$\mathbb{H}_{\left(  0\right)  }^{a}$ and $\mathbb{V}_{\left(  0\right)
}^{ab}$ correspond to leading, $\lambda$-independent terms, and $\mathbb{H}%
_{\left(  p\right)  }^{a}$ and $\mathbb{V}_{\left(  p\right)  }^{ab}$
correspond to deviations of order $\epsilon^{p}$ to the geometric optics.
The leading term $\mathbb{H}_{\left(  0\right)  }^{a}$ defines the
polarization $P^{a}=P^{a}{}_{b}e^{b}\in \mathbb{C}$ and amplitude $\varphi
\in \mathbb{R}$ of the gravitational wave as
\begin{equation}
    \mathbb{H}_{\left(  0\right)  }^{a}=\varphi P^{a}\,.\label{Eq_Def_Ampl_Pol}
\end{equation}

We can always choose to normalize the polarization as
\begin{equation}
    \bar{P}_{ab}P^{ab}=1 \label{Eq_Pol_Norm}
\end{equation}
without losing generality\footnote{The same should be done with $V^{a}{}_{b}$
	when dealing with propagating torsion. In the simple ECSK case, torsion does
	not propagate, and this decomposition of $V^{a}{}_{b}$ as amplitude and
	polarization is unnecessary.}.
Regardless of the theory, it is always possible to use the local Lorentz
invariance and the Lie derivative to prove that the components $H_{ab}$ must
be symmetric, $H_{ab}=H_{ba}$ and to impose the Lorenz gauge on a
Riemann-Cartan geometry as (see \cite{Barrientos:2019awg} for further details)
\begin{equation}
\mathcal{D}_{a}H^{a}-\frac{1}{2}\mathrm{dI}_{a}H^{a}=0\,. \label{loga}
\end{equation}

However, for a generic nonvanishing torsion theory, it is impossible to make
further gauge fixing, and the six possible gravitational polarization modes
could be present in the components of $H_{ab}$. For instance, it is possible
to write the polarization of a gravitational wave propagating in the third
direction of the orthonormal frame as%
\begin{equation}
P_{ab}=p_{\left(  +\right)  }\mathbb{P}_{ab}^{\left(  +\right)  }+p_{\left(
	\times \right)  }\mathbb{P}_{ab}^{\left(  \times \right)  }+p_{\left(  b\right)
}\mathbb{P}_{ab}^{\left(  b\right)  }+p_{\left(  l\right)  }\mathbb{P}%
_{ab}^{\left(  l\right)  }+p_{\left(  x\right)  }\mathbb{P}_{ab}^{\left(
	x\right)  }+p_{\left(  y\right)  }\mathbb{P}_{ab}^{\left(  y\right)  }\,,
\end{equation}
with the orthonormal polarization basis
\begin{align}
\mathbb{P}_{ab}^{\left(  +\right)  } &  =\frac{1}{\sqrt{2}}\left(
\begin{array}
[c]{cccc}%
0 & 0 & 0 & 0\\
0 & 1 & 0 & 0\\
0 & 0 & -1 & 0\\
0 & 0 & 0 & 0
\end{array}
\right)  ,\qquad \mathbb{P}_{ab}^{\left(  \times \right)  }=\frac{1}{\sqrt{2}%
}\left(
\begin{array}
[c]{cccc}%
0 & 0 & 0 & 0\\
0 & 0 & 1 & 0\\
0 & 1 & 0 & 0\\
0 & 0 & 0 & 0
\end{array}
\right) \, ,\\
\mathbb{P}_{ab}^{\left(  b\right)  } &  =\frac{1}{\sqrt{2}}\left(
\begin{array}
[c]{cccc}%
0 & 0 & 0 & 0\\
0 & 1 & 0 & 0\\
0 & 0 & 1 & 0\\
0 & 0 & 0 & 0
\end{array}
\right)  ,\qquad \mathbb{P}_{ab}^{\left(  l\right)  }=\left(
\begin{array}
[c]{cccc}%
0 & 0 & 0 & 0\\
0 & 0 & 0 & 0\\
0 & 0 & 0 & 0\\
0 & 0 & 0 & 1
\end{array}
\right)  \,,\\
\mathbb{P}_{ab}^{\left(  x\right)  } &  =\frac{1}{\sqrt{2}}\left(
\begin{array}
[c]{cccc}%
0 & 0 & 0 & 0\\
0 & 0 & 0 & 1\\
0 & 0 & 0 & 0\\
0 & 1 & 0 & 0
\end{array}
\right) \, ,\qquad \mathbb{P}_{ab}^{\left(  y\right)  }=\frac{1}{\sqrt{2}}\left(
\begin{array}
[c]{cccc}%
0 & 0 & 0 & 0\\
0 & 0 & 0 & 0\\
0 & 0 & 0 & 1\\
0 & 0 & 1 & 0
\end{array}
\right) \, .
\end{align}
and where
\begin{equation}
\bar{p}_{\left(  +\right)  }p_{\left(  +\right)  }+\bar{p}_{\left(
	\times \right)  }p_{\left(  \times \right)  }+\bar{p}_{\left(  b\right)
}p_{\left(  b\right)  }+\bar{p}_{\left(  l\right)  }p_{\left(  l\right)
}+\bar{p}_{\left(  x\right)  }p_{\left(  x\right)  }+\bar{p}_{\left(
	y\right)  }p_{\left(  y\right)  }=1\,.
\end{equation}

It is essential to remember that, even in the case of standard torsionless GR, further gauge fixing (as the transverse traceless gauge) on a generic background geometry is only valid at leading and subleading orders in the eikonal expansion. This way, when we say that the only GW modes in standard GR are $\left(+\right)$ and $\left(\times\right)$, that affirmation is valid only at leading and subleading eikonal orders. This way, higher-order deviations of geometric optics generally include all six polarization modes, but $\left(+\right)$ and $\left(\times\right)$ polarizations vastly dominate over others,

\begin{align}
\bar{p}_{\left(  +\right)  }p_{\left(  +\right)  }+\bar{p}_{\left(
	\times \right)  }p_{\left(  \times \right)  }  & \approx1\,,\label{Eq_hierarch_1}\\
\bar{p}_{\left(  b\right)  }p_{\left(  b\right)  }+\bar{p}_{\left(  l\right)
}p_{\left(  l\right)  }+\bar{p}_{\left(  x\right)  }p_{\left(  x\right)
}+\bar{p}_{\left(  y\right)  }p_{\left(  y\right)  }  & \leq \epsilon^{2}\,.\label{Eq_hierarch_epsilon2}
\end{align}

As we will see in Sec.~\ref{Sec3.2}, torsion can create anomalous polarization propagation. Therefore, the hierarchy of polarization modes Eq.~(\ref{Eq_hierarch_epsilon2}) could break down for GW propagating on a generic torsional background. For this reason, in principle, the modes beyond $\left(+\right)$ and $\left(\times \right)$ could become significant after propagation.


\section{ECSK theory as a playground to study GW propagation}\label{SecIII}

The current article focuses on ECSK theory for several reasons. First of
all, the great observational success of standard GR makes it attractive to
analyze theories close to its dynamics (as ECSK). Second, ECSK is the simplest theory
with nonvanishing torsion, but despite having a non-propagating torsion, it
can affect the propagation of gravitational waves in a non-trivial way.
For these reasons, ECSK provides an excellent playground to study the
phenomenology of GW propagation before jumping to consider more complex
Lagrangians. For other theories, the explicit form of some subleading
couplings will vary (see Sec.~\ref{Sec3.2}), but the general procedure also applies to
those cases.

In terms of the vierbein and the spin connection, the ECSK Lagrangian 4-form
reads%
\begin{equation}
	\mathcal{L}_{\mathrm{EH}}=\frac{1}{\kappa_{\mathrm{4}}}\left(  \frac{1}%
	{4}\epsilon_{abcd}R^{ab}\wedge e^{c}\wedge e^{d}-\frac{\Lambda}{4!}%
	\epsilon_{abcd}e^{a}\wedge e^{b}\wedge e^{c}\wedge e^{d}\right)
	+\mathcal{L}_{\mathrm{M}}\left(  e,\omega,\psi \right)\,,
\end{equation}
where $\mathcal{L}_{\mathrm{M}}$ corresponds to the matter Lagrangian 4-form.
The equations of motion provided by independent variations of the vierbein and
the spin connection correspond to%
\begin{align}
	\frac{1}{2}\epsilon_{abcd}R^{ab}\wedge e^{c}-\frac{\Lambda}{3!}\epsilon_{abcd}e^{a}\wedge e^{b}\wedge e^{c} &  =\kappa_{\mathrm{4}}\ast \tau_{d}\,,\label{deoc1}\\
	\epsilon_{abcd}T^{c}\wedge e^{d} &  =\kappa_{\mathrm{4}}\ast \sigma_{ab}\,,\label{deoc2}
\end{align}
where the variations of $\mathcal{L}_{\mathrm{M}}^{\left(  4\right)  }$ define
the stress-energy 1-form $\tau_{d}=\tau_{nd}e^{n}$ and the spin
density 1-form $\sigma^{ab}=\sigma_{c}{}^{ab}e^{c}$ as
\begin{align}
	\delta_{e}\mathcal{L}_{\mathrm{M}} &  =-\ast \tau_{d}\wedge \delta e^{d}\,, \\
	\delta_{\omega}\mathcal{L}_{\mathrm{M}} &  =-\frac{1}{2}\delta \omega^{ab}\wedge \ast \sigma_{ab}\,. 
\end{align}

It is straightforward to observe that the torsion components are algebraically related to matter's spin tensor, Eq.~(\ref{deoc2}). Therefore, in a vacuum, torsion identically vanishes and cannot propagate. When besides this, we consider the Standard Model as the matter Lagrangian, torsion would seem to be doomed to irrelevance in the context of ECSK. In the Standard Model, only fermions\footnote{In the physics literature, it is trendy to introduce boson-torsion coupling by following the recipe of changing partial derivatives on Minkowski space by covariant derivatives. For instance, the electromagnetic field components become $\nabla_{\mu}A_{\nu}-\nabla_{\nu}A_{\mu}$ with $\nabla=\partial +\Gamma$ through this mechanism, creating an electromagnetism-torsion coupling. However, it is unnecessary and unnatural from a mathematical point of view. Gauge fields correspond to curvatures $F=\mathrm{d}A+\frac{1}{2}\left[ A,A\right]$ on principal bundles~\cite{deAzcarraga:1995jw}, with $\mathrm{d}$ an exterior derivative and nothing else. For instance, from the principal bundle point of view, the electromagnetic field strength corresponds only to the 2-form $F=\mathrm{d}A$. Introducing the affine connection $\Gamma$ in its definition seems like putting fibers on top of the fibers of a fiber bundle.} are a source of torsion (they have a nonvanishing spin tensor) and are affected by it~\cite{Blagojevic:2013xpa,SupergravityVanProeyen}. However, the expected
effect is so weak that there are no realistic particle physics experiments
that could detect torsion in the foreseeable future, and some references even
jokingly advise against betting for a detection of this kind (see the end
Chapter 8 of Ref.~\cite{SupergravityVanProeyen}).
Even more, Standard Model fermions interact forming localized structures.
Since any torsional effect they create cannot propagate in a vacuum, it would
seem that torsion cannot play a role in the late evolution of the
universe\footnote{The situation is contrary when considering extremely dense
	fermion plasma, as in the very early universe or during a black hole collapse.
	In this case, torsion can avoid the singularity, provide an alternative to
	inflation models and give origin to Big Bounce models, see Ref.~\cite{Poplawski:2011jz,Poplawski:2012qy,Cubero:2019lxw, Poplawski:2020hrp, Poplawski:2020kzc}. Regarding
	late universe evolution, it is interesting to observe that including torsional
	effects in vacuum fluctuations reduces the cosmological constant problem from
	122 orders of magnitude to just 8, see Ref.~\cite{Poplawski:2011qr}.}.

However, it is necessary to take into account dark matter before dismissing torsion. It is not clear yet whether dark matter carries spin and we do not know whether its spin tensor vanishes or not. Standard cosmological models generally assume a vanishing spin density for simplicity, but there is no physical reasons behind this hypothesis: the observational constraints on torsion are poor. At the contrary, assuming a nonvanishing dark matter spin density provides a simple explanation for phenomena as the Hubble parameter tension, see~\cite{Izaurieta:2020xpk}.

Of course, having dark matter as a significant source of torsion implies it cannot be any Standard Model-like particle. Spin-1/2 Standard Model particles have a nonvanishing spin tensor, but it is of quartic order in $\psi$ and highly localized. It is straightforward to prove that such a spin tensor cannot create any leading or subleading GW effect. Therefore, the only worth exploring alternative left is acknowledging our ignorance and considering a completely arbitrary dark matter spin tensor instead of the standard one for spin-1/2 particles. However, it opens an intriguing possibility. As we shall see, a torsion background affects GW propagation of its amplitude and polarization. It means that, in principle, a torsion background could affect a gravitational wave propagating over a long cosmological distance. Even worst, it may seem that an unaccounted anomalous propagation of GW amplitude could hinder our efforts to use mergers as standard sirens, making them appear farther or closer than they are.
In the current article, we prove in a model-independent way that it is not the case, and that an unrealistically strong torsion background would be necessary to affect amplitude propagation in an observable way. The following section analyzes GW on an ECSK theory with an arbitrary nonvanishing torsional background.

\subsection{Propagation of GW in ECSK theory}

When studying GW (in any theory), we must separate high-frequency and
low-frequency terms (up to second order in perturbations). The high-frequency
piece describes how the wave propagates on the background geometry, while the
low-frequency piece describes how the wave may affect the background geometry,
creating an effective stress-energy tensor and spin tensor. Chapters 1 and 4 of
Ref.~\cite{Maggiore:1900zz} provide an excellent example of how to perform this separation in the context of GR.

In this article, we will focus only on the propagation of the GW (i.e.,
high-frequency effects). In this high-frequency piece, not all the terms are
equally important. Therefore, this article will consider only terms
contributing to the leading and subleading orders in the eikonal
approximation. 
This way, we start by studying the perturbation of all terms of Eq.(\ref{deoc1}).
However, proceeding as in standard GR (see Ref.~\cite{Barrientos:2019awg} and Eq.~(1.174) of Ref.~\cite{Maggiore:1900zz}), it is straightforward to
prove that from Eq.(\ref{deoc1}) the only perturbation terms contributing to GW
propagation in this approximation are inside of the expression%
\begin{equation}
	\frac{1}{2}\epsilon_{abcn}R_{\left(  1\right)  }^{ab}\wedge e^{c}=0\,,
\end{equation}
where $R_{\left(  1\right)  }^{ab}$ corresponds to the linear perturbations of
Lorentz curvature, Eq.(\ref{deoc1})

It is possible to prove (see Appendix A) of Ref.~\cite{Barrientos:2019awg}) that
\begin{equation}
	\frac{1}{2}\epsilon_{abcn}R_{\left(  1\right)  }^{ab}\wedge e^{c}=\left(
	W_{mn}-\frac{1}{2}\eta_{mn}W^{p}{}_{p}\right)  \ast e^{m}\,,\label{Eq_Mov_High}%
\end{equation}
where after some algebra and using the Lorenz gauge Eq.(\ref{loga}), we have
\begin{equation}
	W_{mn}=-\frac{1}{2}\left(  \mathrm{I}_{n}\mathcal{D}_{a}\mathcal{D}^{a}%
	H_{m}-\mathrm{I}_{n}\left[  \mathcal{D}_{a},\mathcal{D}_{m}\right]
	H^{a}\right)  +\left(  \mathrm{I}_{n}\mathcal{D}_{a}-\mathcal{D}_{n}%
	\mathrm{I}_{a}\right)  V^{a}{}_{m}\,.\label{Eq_Wmn_orig}%
\end{equation}
Therefore, tracing equation (\ref{Eq_Mov_High}) we arrive  (at leading and subleading order in the eikonal approximation, compare with Eqs.~(1.174)-(1.179) of Ref.~\cite{Maggiore:1900zz} for the torsionless case) to 

\begin{equation}
	W_{mn}=0.\label{Eq_W_mn}
\end{equation}

Observe that the wave operator in Eq.~(\ref{Eq_Wmn_orig}) corresponds to the generalized Beltrami
operator, already found in Eq.(\ref{Eq_Weitzenboeck}).

\subsection{Eikonal analysis}\label{Sec3.2}

Up to this point, we have preferred the conciseness of differential forms on
the orthonormal basis to study the general properties of the wave operator and
ECSK theory. However, in the following sections, we must analyze the GW
eikonal limit and their propagation on a cosmological background. It is much
more friendly for most readers to carry out this work in terms of tensors on a
coordinate basis, so we will move to this description in what follows.

Firstly, it is convenient to replace the open superalgebra
Eq.~(\ref{os1})-(\ref{os2}) in Eq.~(\ref{Eq_Wmn_orig}), preserving only leading
($\mathcal{O}\left(  \epsilon^{-2}\right)  $) and subleading ($\mathcal{O}%
\left(  \epsilon^{-1}\right)  $) terms in Eq.~(\ref{Eq_Wmn_orig}) in the eikonal expansion\footnote{In the following, we will closely follow the classical procedure for analyzing the GW eikonal limit, with the key difference being that we will not assume zero torsion. It is important to note that the eikonal expansion is not an expansion in a small parameter like the Taylor series, but rather an expansion used to separate terms that vary on vastly different length scales and therefore have very different characteristic derivative sizes. In some references (see for example Eq. (1.184) of Ref.~\cite{Maggiore:1900zz} and Eq. (35.75) of Ref.~\cite{Misner1973Gravitation}), it is customary to keep the eikonal parameter within the expansion as a fictitious parameter that reminds us of the characteristic size of the term that the derivative acts on. This can be useful as a mnemonic, but we will omit it here to avoid confusing the reader.
} Eqs.~(\ref{Eq_H_eikonal_expan}) and~(\ref{Eq_V_eikonal_expan}), where $\epsilon$ corresponds to the eikonal parameter Eq.~(\ref{Eq_Eikonal_Parameter}). After this, we can move to the coordinate basis using Eq.~(\ref{Eq_D=D}). The result is
\begin{equation}
	\left.  W_{\mu \nu}\right \vert _{\mathrm{lead.}+\mathrm{sublead.}}=-\left[
	\frac{1}{2}\mathcal{\nabla}_{\lambda}\mathcal{\nabla}^{\lambda}H_{\mu \nu
	}+T_{\sigma \rho \nu}\mathcal{\nabla}^{\rho}H^{\sigma}{}_{\mu}+\frac{1}%
	{2}T_{\rho \sigma \mu}\mathcal{\nabla}^{\rho}H^{\sigma}{}_{\nu}\right]
	+\mathcal{\nabla}^{\lambda}V_{\lambda \mu \nu}+\mathcal{\nabla}_{\nu}%
	V_{\mu \lambda}{}^{\lambda}\,.
\end{equation}
This equation has a symmetric and an antisymmetric piece, and from Eq.~(\ref{Eq_W_mn}) it is clear they must
independently vanish,
\begin{align}
	\left.  W_{\mu \nu}^{+}\right \vert _{\mathrm{lead.}+\mathrm{sublead.}}  &
	=0\,,\\
	\left.  W_{\mu \nu}^{-}\right \vert _{\mathrm{lead.}+\mathrm{sublead.}}  &  =0\,.
\end{align}
The vanishing of $W_{\mu \nu}^{-}$ allow us to solve $V_{\alpha \beta \gamma}$ in
terms of $H_{\mu \nu}$ and $T_{\lambda \mu \nu}$ as%
\begin{align}
	V_{\alpha \beta \gamma}=&\frac{1}{4}\left(  \left[  T^{\sigma}{}_{\sigma \rho
	}H^{\rho}{}_{\alpha}-T_{\rho \sigma \alpha}H^{\rho \sigma}\right]  g_{\beta
		\gamma}-\left[  T^{\sigma}{}_{\sigma \rho}H^{\rho}{}_{\beta}-T_{\rho \sigma
		\beta}H^{\rho \sigma}\right]  g_{\alpha \gamma}\right)+ \\
	&-\left(  T_{\rho
		\alpha \beta}+\frac{1}{4}\left[  T_{\alpha \beta \rho}-T_{\beta \alpha \rho
	}\right]  \right)  H^{\rho}{}_{\gamma}+\frac{1}{4}\left[  \left(  T_{\beta
		\rho \gamma}+T_{\gamma \rho \beta}\right)  H^{\rho}{}_{\alpha}-\left(
	T_{\alpha \rho \gamma}+T_{\gamma \rho \alpha}\right)  H^{\rho}{}_{\beta}\right]\,. \nonumber
\end{align}
Here we can see that a vacuum $T_{\lambda \mu \nu}=0$ also implies
$V_{\alpha \beta \gamma}=0$, as it should be expected for non-propagating torsion.
Replacing it back in $\left.  W_{\mu \nu}^{+}\right \vert _{\mathrm{lead.}%
	+\mathrm{sublead.}}$, we obtain%
\begin{align}
	\left.  W_{\mu\nu}^{+}\right\vert _{\mathrm{lead.+sublead}}  & =\frac{1}%
	{2}\left[  -\nabla_{\lambda}\nabla^{\lambda}H_{\mu\nu}-\left(  2T_{\sigma
		\rho\nu}+\frac{1}{2}\left[  T_{\rho\sigma\nu}+T_{\nu\sigma\rho}\right]
	\right)  \nabla^{\rho}H_{~\mu}^{\sigma}\right.  +\\
	& \left.  -\left(  2T_{\sigma\rho\mu}+\frac{1}{2}\left[  T_{\sigma\rho\mu
	}+T_{\mu\sigma\rho}\right]  \right)  \nabla^{\rho}H_{~\nu}^{\sigma}-\frac
	{1}{2}g_{\mu\nu}T_{\rho\sigma\lambda}\nabla^{\lambda}H^{\rho\sigma}+\frac
	{1}{4}\left[  g_{\mu\nu}T_{~\sigma\lambda}^{\sigma}-\left(  T_{\mu\nu\lambda
	}+T_{\nu\mu\lambda}\right)  \right]  \partial^{\lambda}H\right] \nonumber
\end{align}

From this expression, it is clear that torsion affects the propagation of GW at
subleading order, and when torsion vanishes, we recover the standard GR wave equation in terms of the standard Beltrami operator.

At leading order $\mathcal{O}\left(  \epsilon^{-2}\right)  $, the dispersion
relation remains unchanged,%
\begin{equation}
	k^{\mu}k_{\mu}=0\,, \label{Eq_Dispersion}%
\end{equation}
implying GW propagation at the speed of light. Even more, taking into account
that $k_{\mu}=\partial_{\mu}\theta$ and taking the derivative of the
dispersion relation, we have that%
\begin{equation}
	k^{\mu}\mathring{\nabla}_{\mu}k^{\lambda}=0\,. \label{Eq_GW_Geodesic}%
\end{equation}
These relations reveal an essential fact. Despite the nonvanishing
background torsion, GW move along null,  torsionless geodesics. It
may seem counterintuitive because, besides standard torsionless geodesics, a
Riemann-Cartan geometry allows defining auto-parallels. Auto-parallels are
curves given by%
\begin{equation}
	\frac{\mathrm{d}^{2}X^{\lambda}}{\mathrm{d}\tau^{2}}+\Gamma_{\mu \nu}^{\lambda
	}\frac{\mathrm{d}X^{\mu}}{\mathrm{d}\tau}\frac{\mathrm{d}X^{\nu}}%
	{\mathrm{d}\tau}=0\,,
\end{equation}
using the full connection $\Gamma_{\mu \nu}^{\lambda}$ instead of only the
Christoffel piece $\mathring{\Gamma}_{\mu \nu}^{\lambda}$.
This simple fact is crucial from an observational point of view. The
multimessenger observation GW170817/GRB170817 implies that GW and
electromagnetic waves travel at the same speed, and on the same kind of
trayectories. If electromagnetic waves moved on null torsionless geodesics and
GW on null auto-parallels, it still could have caused a delay even if both
waves traveled at the same speed.

The subleading order $\mathcal{O}\left(  \epsilon^{-1}\right)  $ provide us
with the relation
\begin{equation}
	\frac{1}{2}\nabla_{\lambda}k^{\lambda}\mathbb{H}_{\mu \nu}+k^{\lambda}\left(
	\nabla_{\lambda}\mathbb{H}_{\mu \nu}+\frac{1}{2}M_{\lambda \mu \nu}^{+}\right)
	=0\,, \label{Ec_Origen_Amplitud_Polarizacion}%
\end{equation}
with%
\begin{align}
	M_{\lambda \mu \nu}^{+}  &  =\left(  2T_{\sigma \lambda \nu}+\frac{1}{2}\left[
	T_{\lambda \sigma \nu}+T_{\nu \sigma \lambda}\right]  \right)  \mathbb{H}^{\sigma
	}{}_{\mu}+\left(  2T_{\sigma \lambda \mu}+\frac{1}{2}\left[  T_{\lambda \sigma
		\mu}+T_{\mu \sigma \lambda}\right]  \right)  \mathbb{H}^{\sigma}{}_{\nu
	}+\label{Ec_Def_M}\\
	&  +\frac{1}{2}g_{\mu \nu}T_{\rho \sigma \lambda}\mathbb{H}^{\rho \sigma}-\frac
	{1}{4}\left[  g_{\mu \nu}T^{\sigma}{}_{\sigma \lambda}-\left(  T_{\mu \nu \lambda
	}+T_{\nu \mu \lambda}\right)  \right]  \mathbb{H}\,.
\end{align}
Notice that Eq.~(\ref{Ec_Origen_Amplitud_Polarizacion}) codifies the amplitude
$\varphi$ and polarization $P_{\mu \nu}$ propagation Eq.~(\ref{Eq_Def_Ampl_Pol}). The time-honored way of getting this information is by defining the `number of rays' current density
$J_{\mu}=\varphi^{2}k_{\mu}$ (also known as `number of photons' current
density in standard optics). In the standard Riemannian case $J_{\mu}$ is
conserved, but it is no longer the case for nonvanishing torsion. In our case,
after replacing Eq.~(\ref{Eq_Def_Ampl_Pol}) and Eq.~(\ref{Eq_Pol_Norm}) in Eq.~(\ref{Ec_Origen_Amplitud_Polarizacion}) (see Appendix \ref{appA}), we get
\begin{equation}
	\mathring{\nabla}_{\lambda}J^{\lambda}=\left(  \Pi^{\mu \nu}-g^{\mu \nu}\right)
	T_{\mu \nu \lambda}J^{\lambda}\,, \label{Eq_Non_Consv_J}%
\end{equation}
with $\Pi^{\mu \nu}$ given by%
\begin{equation}
	\Pi^{\mu \nu}=\frac{1}{2}\left[  3\left(  \bar{P}^{\mu \sigma}P_{\sigma}{}^{\nu
	}+P^{\mu \sigma}\bar{P}_{\sigma}{}^{\nu}\right)  -\left(  \bar{P}P^{\mu \nu
	}+\bar{P}^{\mu \nu}P\right)  +\frac{1}{2}\bar{P}Pg^{\mu \nu}\right]\,,
	\label{Eq_Pi_mn}%
\end{equation}
and $P=P^{\lambda}{}_{\lambda}$ denoting the trace of the polarization.

The polarization also propagates anomalously on a torsional background.
Replacing Eq.(\ref{Eq_Non_Consv_J}) in
Eq.~(\ref{Ec_Origen_Amplitud_Polarizacion}), we find after some algebra (see
Appendix \ref{appA}) the relation
\begin{align}
	k^{\lambda}\mathring{\nabla}_{\lambda}P_{\mu \nu} &  =-\frac{1}{2}k^{\lambda
	}\left[  \Pi^{\rho \sigma}T_{\rho \sigma \lambda}P_{\mu \nu}+\left(
	T_{\sigma \lambda \nu}-\frac{1}{2}\left[  T_{\lambda \sigma \nu}+T_{\nu
		\sigma \lambda}\right]  \right)  P^{\sigma}{}_{\mu}+\right.
	\label{Eq_Polarization_Propag}\\
	&  \left.  +\left(  T_{\sigma \lambda \mu}-\frac{1}{2}\left[  T_{\lambda
		\sigma \mu}+T_{\mu \sigma \lambda}\right]  \right)  P^{\sigma}{}_{\nu}+\frac
	{1}{2}g_{\mu \nu}T_{\rho \sigma \lambda}P^{\rho \sigma}-\frac{1}{4}\left[
	g_{\mu \nu}T^{\sigma}{}_{\sigma \lambda}-\left(  T_{\mu \nu \lambda}+T_{\nu
		\mu \lambda}\right)  \right]  P\right]  .\nonumber
\end{align}
Equations (\ref{Eq_Non_Consv_J}) and (\ref{Eq_Polarization_Propag}) show
that torsion, even if it does not propagate, can affect the propagation of the GW
amplitude and polarization.

Some comments are in order. Regarding Eq.~(\ref{Eq_Non_Consv_J})-(\ref{Eq_Polarization_Propag}), the explicit form of the right-hand side of
Eq.~(\ref{Eq_Non_Consv_J})-(\ref{Eq_Polarization_Propag}) depends on the ECSK
Lagrangian structure. However, it is a generic feature of theories with
nonvanishing torsion to create an anomalous propagation of amplitude and
polarization. This anomalous propagation is an intrinsic feature of the wave
operator on Riemann-Cartan geometries, and changing of theory would only
change coefficients weights in Eq.~(\ref{Eq_Non_Consv_J})-(\ref{Eq_Polarization_Propag}). For instance, even if we start with just the
operator
\begin{equation}
	\mathcal{D}_{a}\mathcal{D}^{a}H_{m}=0
\end{equation}
instead of vanishing the whole Eq.~(\ref{Eq_Wmn_orig}), it will still lead to
anomalous propagation, see Ref.~\cite{Barrientos:2019msu}.

To have a better understanding of how torsion causes the anomalous propagation
of amplitude and polarization, it is practical to split torsion in different
components, and to study the consequences of each one of them.

In particular, in $d=4$ it is always possible to decompose the torsion tensor
as%
\begin{equation}
	T_{\lambda \mu \nu}=\frac{2}{3}\left(  \mathcal{T}_{\lambda \mu \nu}%
	-\mathcal{T}_{\lambda \nu \mu}\right)  +\frac{1}{3}\left(  \eta_{\lambda \mu
	}\mathcal{V}_{\nu}-\eta_{\lambda \nu}\mathcal{V}_{\mu}\right)  +\sqrt
	{\left \vert g\right \vert }\epsilon_{\lambda \mu \nu \rho}\mathcal{A}^{\rho},
\end{equation}
where $\mathcal{A}^{\rho}$ corresponds to the axial component, $\mathcal{V}%
_{\mu}$ to the vectorial component, and $\mathcal{T}_{\lambda \mu \nu}$ to the
tensorial one (see Ref.~\cite{Aldrovandi:2013}). The tensorial components
satisfy%
\begin{align}
	\mathcal{T}_{\lambda \mu \nu}  & =\mathcal{T}_{\mu \lambda \nu}\,,\\
	\mathcal{T}^{\lambda}{}_{\lambda \nu}  & =0\,,\\
	\mathcal{T}_{\lambda \mu}{}^{\mu}  & =0\,,\\
	\mathcal{T}_{\lambda \mu \nu}+\mathcal{T}_{\nu \lambda \mu}+\mathcal{T}_{\mu
		\nu \lambda}  & =0\,.
\end{align}
In terms of these components, after some algebra, it is possible to prove that
the anomalous propagation of amplitude, Eq.~(\ref{Eq_Non_Consv_J}), reduces to%
\begin{equation}
	\mathring{\nabla}_{\lambda}J^{\lambda}=\left(  \Pi^{ab}\mathcal{T}_{abc}%
	-\frac{1}{3}\Pi_{bc}\mathcal{V}^{b}\right)  J^{c}.
\end{equation}
Following a similar treatment, the anomalous propagation of polarization, Eq.~(\ref{Eq_Polarization_Propag}), reduces to
\begin{equation}
	k^{\lambda}\mathring{\nabla}_{\lambda}P_{\mu \nu}=-\frac{1}{2}k^{\lambda
	}\left(  I_{\mu \nu \lambda}^{\left(  T\right)  }+I_{\mu \nu \lambda}^{\left(
		V\right)  }+I_{\mu \nu \lambda}^{\left(  A\right)  }\right)  ,
\end{equation}~
where the tensorial, vectorial, and axial contributions to the anomalous
propagation of polarization correspond to%
\begin{align}
	I_{\mu \nu \lambda}^{\left(  T\right)  } &  =\mathcal{T}_{\rho \sigma \lambda
	}\left(  \Pi^{\rho \sigma}P_{\mu \nu}+\frac{1}{2}\eta_{\mu \nu}P^{\rho \sigma
	}\right)  +\frac{1}{2}\mathcal{T}_{\mu \nu \lambda}P+\frac{1}{3}\left(
	\mathcal{T}_{\rho \lambda \nu}-3\mathcal{T}_{\nu \rho \lambda}+2\mathcal{T}%
	_{\lambda \nu \rho}\right)  P^{\rho}{}_{\mu}+\\
	&  +\frac{1}{3}\left(  \mathcal{T}_{\rho \lambda \mu}-3\mathcal{T}_{\mu
		\rho \lambda}+2\mathcal{T}_{\lambda \mu \rho}\right)  P^{\rho}{}_{\nu},\\
	I_{\mu \nu \lambda}^{\left(  V\right)  } &  =\frac{1}{3!}\mathcal{V}^{\sigma
	}\left[  2\left(  3\eta_{\lambda \sigma}-\Pi_{\lambda \sigma}\right)  P_{\mu \nu
	}-\eta_{\mu \nu}P_{\lambda \sigma}+\left(  \eta_{\rho \lambda}\eta_{\nu \sigma
	}-3\eta_{\nu \rho}\eta_{\lambda \sigma}+2\eta_{\lambda \nu}\eta_{\rho \sigma
	}\right)  P^{\rho}{}_{\mu}+\right.  \\
	&  \left.  +\left(  \eta_{\rho \lambda}\eta_{\mu \sigma}-3\eta_{\mu \rho}%
	\eta_{\lambda \sigma}+2\eta_{\lambda \mu}\eta_{\rho \sigma}\right)  P^{\rho}%
	{}_{\nu}+\frac{1}{2}\left(  \eta_{\mu \nu}\eta_{\lambda \sigma}-\eta_{\mu
		\lambda}\eta_{\nu \sigma}-\eta_{\nu \lambda}\eta_{\mu \sigma}\right)  P\right]
	,\\
	I_{\mu \nu \lambda}^{\left(  A\right)  } &  =\sqrt{\left \vert g\right \vert
	}\mathcal{A}^{\sigma}\left(  \epsilon_{\nu \rho \lambda \sigma}P^{\rho}{}_{\mu
	}+\epsilon_{\mu \rho \lambda \sigma}P^{\rho}{}_{\nu}\right)  .
\end{align}

From here, the main result is that only the tensorial and vectorial components
of torsion contribute to the anomalous propagation of amplitude, and the axial
components do not. However, all the components of torsion contribute to the
anomalous propagation of polarizations, including the axial ones. This fact could be important when considering spin-1/2 particles plasma as a source of torsion (their spin tensor is axial) but it is not the case we are analysing here.

In the following sections, we will focus on propagation on cosmological
scales. The Copernican symmetries are compatible only with the vectorial and
axial mode, simplifying the analysis.

\subsection{Propagation of amplitude in a weak torsion scenario}

When considering gravitational waves emitted by a binary system, torsion can potentially have effects in two distinct stages. Firstly, during the emission process itself, due to the fact that the sources (e.g. neutron stars) have a non-trivial spin tensor within them due to their matter content. The second possibility is that the medium through which gravitational waves propagate has non-zero torsion; this would be the case for example if dark matter possessed a non-trivial spin density. In that case, torsion could in principle cumulatively affect the GW as it travels over a cosmological distance. 

The first case (torsional effects due to the non-trivial spin tensor of the sources) was studied in Refs.~\cite{Battista:2021rlh,Battista:2022hmv,Battista:2022sci} within the context of the ECSK theory to the first post-Newtonian order using the Damour-Iyer-Banchet formulation. It is possible to observe there that in this case, the potential torsional effects are extremely small and difficult to distinguish from standard GR, in line with what current observations indicate (observations in concurrence with standard General Relativity, falling within the margin of experimental error). To detect significant deviations in the emission process, it is imperative (i) to be in close proximity to the merger or alternatively, (ii) to consider a background with exceptionally high spin densities, such as those present in the case of primordial gravitational waves in the early universe. Therefore, for current observations (far late universe mergers) deviations in emission are expected to be tiny.

For this reason, in this article, we will focus on the second case (potential cumulative torsional effects when propagating over a long distance), and we will consider as initial conditions what we will call a \textquotedblleft weak torsion scenario\textquotedblright  for the emission process, i.e. that a torsion background may be present but is too weak to affect the emission process of the gravitational wave in an observable manner.
%
%
In particular, we will consider that at the emission moment only the polarization modes $\left(  +\right)  $ and $\left(  \times \right)  $ are relevant, and the other four modes are at least $\epsilon$ times weaker (see Eq.~\ref{Eq_hierarch_epsilon2}), as it happens in standard GR. However, the GW has to travel over a long cosmological distance before reaching our detectors. In such case, it is non-trivial to decide whether or not the cumulative effect of the torsional background Eq.~(\ref{Eq_Non_Consv_J})-(\ref{Eq_Polarization_Propag}) could have observational consequences at the moment of detection.

In particular, the scrambling of polarization modes along the geodesic
Eq.~(\ref{Eq_Polarization_Propag}) could make grow the other polarization
modes while the GW propagates. Furthermore, an unaccounted anomalous propagation of GW
amplitude Eq.~(\ref{Eq_Non_Consv_J}) could hinder our efforts to use mergers
as standard sirens, making them appear farther or closer than they are.
In this article we focus our attention in the amplitude problem; the details
of polarization propagation are left for future work.

Let us consider a GW propagating along the null torsionless geodesic
Eq.~(\ref{Eq_GW_Geodesic}), with a $\frac{\mathrm{d}X^{\mu}}{\mathrm{d}\eta
}\propto k^{\mu}$ tangent vector and an affine parameter $\eta$, and let us
call $\eta_{0}$ the affine parameter at the moment of emission.
Let $\varphi \left(  \eta \right)  $ be the amplitude of the gravitational
wave at $\eta,$ and $\mathring{\varphi}\left(  \eta \right)  $ the amplitude
predicted at $\eta$ by standard GR. Then, we can define%
\begin{equation}
	A\left(  \eta \right)  =\ln \frac{\varphi}{\mathring{\varphi}}\,,%
\end{equation}
as a parameter describing the anomalous propagation of amplitude. The weak
torsion scenario implies that $\varphi \left(  \eta_{0}\right)  =\mathring
{\varphi}\left(  \eta_{0}\right)$, and therefore $A\left(  \eta_{0}\right)  =0$\,. In this context, let us assume a generic theory with non-vanishing torsion, which lead us to a breaking of the conservation of $J_{\mu}$ in some generalized
form of Eq.~(\ref{Eq_Non_Consv_J}),%
\begin{equation}
	\mathring{\nabla}_{\mu}J^{\mu}=N_{\mu}J^{\mu}\,.
\end{equation}
Since $\varphi=\mathrm{e}^{A}\mathring{\varphi}$, it is possible to write this
last equation as%
\begin{equation}
	\partial_{\mu}\mathrm{e}^{2A}\mathring{\varphi}^{2}k^{\mu}+\mathrm{e}%
	^{2A}\mathring{\nabla}_{\mu}\mathring{J}^{\mu}=N_{\mu}\mathrm{e}^{2A}%
	\mathring{\varphi}^{2}k^{\mu}\,,
\end{equation}
where $\mathring{J}^{\mu}=\mathring{\varphi}^{2}k^{\mu}$. Furthermore, using the fact that $\mathring
{\nabla}_{\mu}\mathring{J}^{\mu}=0$\,, after a bit of algebra one finds%
\begin{equation}
	\frac{\mathrm{d}A}{\mathrm{d}\eta}=\frac{1}{2}N_{\mu}\frac{\mathrm{d}X^{\mu}%
	}{\mathrm{d}\eta}\,.
\end{equation}
In this way, we have that $A=\frac{1}{2}\int_{\eta_{0}}^{\eta}\mathrm{d}%
\tilde{\eta}\,N_{\mu}\left(  \tilde{\eta}\right)  \frac{\mathrm{d}X^{\mu}%
}{\mathrm{d}\tilde{\eta}}$ and
\begin{equation}
	\varphi \left(  \eta \right)  =\mathrm{e}^{\frac{1}{2}\int_{\eta_{0}}^{\eta
		}\mathrm{d}\tilde{\eta}\,N_{\lambda}\left(  \tilde{\eta}\right)
		\frac{\mathrm{d}X^{\lambda}}{\mathrm{d}\tilde{\eta}}}\mathring{\varphi}\left(
	\eta \right)\,,\label{Eq_phi vs phi}
\end{equation}
where
\begin{equation}
	N_{\lambda}=\left(  \Pi^{\mu \nu}-g^{\mu \nu}\right)  T_{\mu \nu \lambda}\,,
\end{equation}
in the case of ECSK theory. Consequently, in the case of mergers, the mismatch between $\varphi \left(  \eta \right)  $
and $\mathring{\varphi}\left(  \eta \right)  $ could lead in principle to a
wrong assessment of the luminosity distance relation.

To be more precise, let us observe that the predicted amplitude for mergers in
standard GR has the form
\begin{equation}
	\mathring{\varphi}=\frac{1}{D_{\mathrm{L}}}F_{\mathrm{GW}}\,,
\end{equation}
where $D_{\mathrm{L}}$ is the luminosity distance, and $F_{\mathrm{GW}}$ is a
complicated, time-dependent function of the masses and angular momentum of the
merger (see chap.~4 of Ref.~\cite{Maggiore:1900zz}).

A torsional background introduces the extra correction factor $\mathrm{e}^{A}$
to the amplitude. Therefore, an observer modeling the GW profile by standard
GR, would assign the wrong luminosity distance $\mathring{D}_{\mathrm{L}}$ to
the source
\begin{equation}
	\varphi=\frac{1}{\mathring{D}_{\mathrm{L}}}F_{\mathrm{GW}}=\mathrm{e}^{A}%
	\frac{1}{D_{\mathrm{L}}}F_{\mathrm{GW}}\,,
\end{equation}
i.e.,
\begin{equation}
	\frac{D_{\mathrm{L}}}{\mathring{D}_{\mathrm{L}}}=\mathrm{e}^{A}\,.\label{Eq_D/D}
\end{equation}
In the weak torsion scenario we consider here, the effect would be noticeable only
when $\left \vert A\right \vert >0$, i.e., when the GW propagates over long
cosmological distances. For this reason and, in order to constrain any possible observable effect, in the following section we will
integrate out these relations, considering different cosmological scenarios involving non-vanishing torsion.

\section{Torsion and cosmological symmetries}\label{SecIV}

A Riemann-Cartan geometry satisfying the Copernican symmetries of homogeneity
and isotropy on a flat spatial section has a metric and torsion tensor of
the form
\begin{align}
\mathrm{d}s^{2}  &  =-c^{2}\mathrm{d}t^{2}+a^{2}\left(  t\right)  \left(
\mathrm{d}x^{2}+\mathrm{d}y^{2}+\mathrm{d}z^{2}\right)\,,\label{Eq_FRW_metric}%
\\
T_{\mu \nu \lambda}  &  =-\frac{1}{c^{2}}\left[  \nu_{+}\left(  t\right)
\left(  g_{\mu \lambda}g_{\nu \rho}-g_{\mu \nu}g_{\lambda \rho}\right)
+2\sqrt{\left \vert g\right \vert }\nu_{-}\left(  t\right)  \epsilon_{\lambda
\mu \nu \rho}\right]  U^{\rho}\,. \label{Eq_FRW_Torsion}%
\end{align}
This way, to describe this geometry, we need the two functions $\nu_{+}\left(
t\right)  $ and $\nu_{-}\left(  t\right)  $ besides the common scale factor
$a\left(  t\right)  $. The symbols $\pm$ refer to the parity of the associated
torsion component.
Using the ansatz Eq.~(\ref{Eq_FRW_Torsion}), one finds%
\begin{align}
N_{\lambda}  &  =\left(  \Pi^{\mu \nu}-g^{\mu \nu}\right)  T_{\mu \nu \lambda}\,,\\
&  =-\frac{1}{c}\nu^{+}\left(  t\right)  \left(  \Pi_{\lambda0}-g_{\lambda
0}-\left(  \Pi^{\sigma}{}_{\sigma}-4\right)  g_{\lambda0}\right)\,.
\end{align}
From Eq.~(\ref{Eq_FRW_Torsion}), we have that $\Pi^{\sigma}{}_{\sigma}=3$, and
therefore%
\begin{equation}
N_{\lambda}=-\frac{1}{4c}\nu^{+}\left(  t\right)  \bar{P}Pg_{\lambda0}\,.
\end{equation}
From this expression for $N_{\lambda}$, we have that
\begin{equation}
A\left(  t\right)  =\frac{1}{4}\int_{t_{0}}^{t}\mathrm{d}\tilde{t}\, \nu
^{+}\left(  \tilde{t}\right)  \frac{1}{2}\left \vert P\right \vert ^{2}\,.
\label{Eq_A(P)}%
\end{equation}
In this way, to solve $A\left(  t\right)  $ we need $P\left(  t\right)  $.
Tracing Eq.~(\ref{Eq_Polarization_Propag}) we get%
\begin{equation}
k^{\lambda}\mathring{\nabla}_{\lambda}P=\frac{1}{2c}\nu^{+}\left(  t\right)
k^{\lambda}\left(  \frac{1}{4}\bar{P}P-\frac{1}{2}\right)  Pg_{\lambda0}\,,
\end{equation}
and since along the geodesic $k^{\lambda}\propto \frac{\mathrm{d}X^{\lambda}}{\mathrm{d}\eta}$,%
\begin{equation}
\frac{\mathrm{d}P}{\mathrm{d}t}=-\frac{1}{2}\nu^{+}\left(  t\right)  \left(
\frac{1}{2}\bar{P}P-1\right)  \frac{1}{2}P\,.
\end{equation}
From here, we have that%
\begin{equation}
\frac{\mathrm{d}}{\mathrm{d}t}\left(  \frac{1}{2}\left \vert P\right \vert
^{2}\right)  =-\frac{1}{2}\nu^{+}\left(  t\right)  \left(  \frac{1}%
{2}\left \vert P\right \vert ^{2}-1\right)  \frac{1}{2}\left \vert P\right \vert
^{2}\,.%
\end{equation}
It is simple to integrate this equation as%
\begin{equation}
\frac{1}{2}\left \vert P\right \vert ^{2}\left(  t\right)  =\frac{1}{\left(
\frac{1}{\frac{1}{2}\left \vert P\right \vert ^{2}\left(  t_{0}\right)
}-1\right)  \mathrm{e}^{-\frac{1}{2}\Theta^{+}}+1}\,,
\end{equation}
with
\begin{equation}
    \Theta^{+}=\int_{t_{0}}^{t}\mathrm{d}\tilde{t}\, \nu^{+}\left(  \tilde{t}\right)\,.\label{Eq_Theta+}
\end{equation}
With this expression, one integrates Eq.~(\ref{Eq_A(P)}) as%
\begin{align}
\mathrm{e}^{A\left(  t\right)  }  &  =\sqrt{\frac{1-\frac{1}{2}\left \vert
P\right \vert ^{2}\left(  t_{0}\right)  }{1-\frac{1}{2}\left \vert P\right \vert
^{2}\left(  t\right)  }},\\
&  =\sqrt{1+\frac{1}{2}\left \vert P\right \vert ^{2}\left(  t_{0}\right)
\left[  \mathrm{e}^{\frac{1}{2}\Theta^{+}}-1\right]  }.\label{Eq_eA_Pchico}
\end{align}

Some comments are in order. First, the anomaly in amplitude propagation only
depends on the $\nu_{+}$ component and not on $\nu_{-}$. It is not strange
when we consider the role of torsion in cosmic evolution; see sec.~\ref{SecIV}. In the
weak torsion scenario, $\frac{1}{2}\left \vert P\right \vert
^{2}\left(  t_{0}\right)\sim\epsilon^{2}  $ is a tiny positive number (see Eq.~\ref{Eq_hierarch_epsilon2})), since in astrophysical situations $\epsilon \leq10^{-20}$. Therefore, the only chance to have an observable deviation of GR is if $\Theta^{+}\gg0$ and big
enough to compensate for the smallness of $\frac{1}{2}\left \vert P\right \vert
^{2}\left(  t_{0}\right)  $. A $\Theta^{+}\le 0$ stands no chance of producing an observable effect in $\mathrm{e}^{A\left(  t\right)  }$. It means that, in principle, ECSK gravity could make mergers to appear closer than they really are ($\mathrm{e}^{A\left(  t\right)  }>1$ in Eq.~(\ref{Eq_D/D})) but not further than they are.

However, when using mergers as standard candles, there will always be some
uncertainty $\delta \mathring{D}_{\mathrm{L}}$ in determining the luminosity
distance $\mathring{D}_{\mathrm{L}}$, see the Ref.~\cite{Holz:2005df}. Therefore, to have
an observable torsion anomaly in the determination of luminosity distances,
it is necessary that $D_{\mathrm{L}}>\mathring{D}_{\mathrm{L}}+\delta 
\mathring{D}_{\mathrm{L}}$, i.e., that 
\begin{equation}
\mathrm{e}^{A}>1+\frac{\delta \mathring{D}_{\mathrm{L}}}{\mathring{D}_{%
\mathrm{L}}}.
\end{equation}%
Considering Eq.~(\ref{Eq_eA_Pchico}), we have 
\begin{equation}
\sqrt{1+\frac{1}{2}\left\vert P\right\vert ^{2}\left( t_{0}\right) \left[ 
\mathrm{e}^{\frac{1}{2}\Theta ^{+}}-1\right] }>1+\frac{\delta \mathring{D}_{%
\mathrm{L}}}{\mathring{D}_{\mathrm{L}}},
\end{equation}%
and therefore%
\begin{equation}
\Theta ^{+}>2\ln \left[ \frac{\left( 1+\frac{\delta \mathring{D}_{\mathrm{L}}%
}{\mathring{D}_{\mathrm{L}}}\right) ^{2}-1}{\frac{1}{2}\left\vert
P\right\vert ^{2}\left( t_{0}\right) }+1\right] .
\end{equation}
Since at most $\frac{1}{2}\left\vert P\right\vert ^{2}\left( t_{0}\right)
\sim \epsilon ^{2}$, the minimal $\Theta _{\min }^{+}$ that could produce a
torsional anomaly above the detection threshold corresponds to 
\begin{equation}
\Theta _{\min }^{+}\approx 2\ln \left[ \left( \frac{\epsilon }{2}\right)
^{-2}\frac{\delta \mathring{D}_{\mathrm{L}}}{\mathring{D}_{\mathrm{L}}}%
\right] .
\end{equation}

Using the performance estimates for LISA (see Ref.~\cite{Holz:2005df}), we have that in
the best scenario 
\begin{equation}
\frac{\delta \mathring{D}_{\mathrm{L}}}{\mathring{D}_{\mathrm{L}}}\sim
10^{-3},
\end{equation}%
(which would be an astounding feat). Considering that $\epsilon <10^{-20}$,
we have that 
\begin{equation}
\Theta _{\min }^{+}\sim 170.\label{Eq_Theta_MIN}
\end{equation}
As we shall see in the following sections, meeting these conditions seems
physically unfeasible for realistic ECSK models: reaching the minimal
detectable value of $\Theta _{\min }^{+}$ would require, to the best of our
knowledge, conditions that cosmological observations rule out. For this
reason, we can conclude that mergers are reliable standard sirens even if
there are unaccounted torsional ECSK effects at play. The mergers' anomalous
amplitude propagation that ECSK torsion could create falls below the LISA
detection threshold.


\subsection{Evaluating $\Theta^{+}$ for different ECSK cosmological models}

It is possible to express $\Theta^{+}$ in terms of the redshift $z=\frac
{a_{0}}{a}-1$ considering that%
\begin{equation}
\mathrm{d}t=-\frac{\mathrm{d}z}{\left(  z+1\right)  H}\,,
\end{equation}
where $H=\frac{\dot{a}}{a}$ corresponds to the Hubble parameter. For a merger at redshift $z$  Eq.~(\ref{Eq_Theta+}) becomes
\begin{equation}
    \Theta^{+}\left(  z\right)  =-\int_{z}^{0}\mathrm{d}\tilde{z}\, \frac{\nu
^{+}\left(  \tilde{z}\right)  }{\left(  \tilde{z}+1\right)  H\left(  \tilde
{z}\right)  }\,.\label{Eq_Theta+Z}
\end{equation}
For $z<1$ it is possible to attempt an estimation of this integral through
observational data; for $z>1$ we can use ECSK cosmological models that agree
with observations.

\subsection{Estimation for merger at $z<1$}

For $z<1,$ we have%
\begin{equation}
\frac{1}{H\left(  z\right)  }=\frac{1}{H_{0}}\left[  1-z\left(  1+q_{0}%
\right)  \right]  +\cdots,
\end{equation}
where $q_{0}$ corresponds to the deceleration parameter. In terms of this
expression, we have%
\begin{equation}
\Theta^{+}\left(  z\right)  \approx-\frac{1}{H_{0}}\int_{z}^{0}\mathrm{d}%
\tilde{z}\, \frac{\nu^{+}\left(  \tilde{z}\right)  }{\tilde{z}+1}\left[
1-z\left(  1+q_{0}\right)  \right]  \,.
\end{equation}
To integrate this last equation, in rigor we need a particular cosmological model of torsion
for $\nu^{+}\left(  \tilde{z}\right)  $. However, for an estimation we will
just introduce some \textquotedblleft representative value\textquotedblright%
\ $\left \langle \nu^{+}\right \rangle $,%
\begin{align}
\Theta^{+}\left(  z\right)   &  \approx-\frac{\left \langle \nu^{+}%
\right \rangle }{H_{0}}\int_{z}^{0}\mathrm{d}\tilde{z}\, \frac{1-\left(
1+q_{0}\right)  z}{1+\tilde{z}}\,,\\
&  =-\frac{\left \langle \nu^{+}\right \rangle }{H_{0}}\left[  \left(
1+q_{0}\right)  z-\left(  2+q_{0}\right)  \ln \left(  1+z\right)  \right] \,.
\end{align}
Using the observational estimations of $q_{0}\sim -0.6$, we conclude that
when $z<1$ then 
\begin{equation}
\Theta ^{+}\left( z\right) <\frac{\left\langle \nu ^{+}\right\rangle }{H_{0}}%
0.6\,,
\end{equation}%
and therefore, to reach the minimal value of eq~.(\ref{Eq_Theta_MIN}) it is necessary
something as%
\begin{equation}
\left\langle \nu ^{+}\right\rangle \sim 100H_{0}\,.
\end{equation}

The estimates for $\nu _{0}^{+}$ in the literature are model-dependent, but $%
\nu _{0}^{+}=100H_{0}$ is orders of magnitude above the even most optimistic
estimates for ECSK cosmology, see Ref.~\cite{Kranas:2018jdc,Pereira:2019yhu,Pereira:2022cmu}. In the following sections we will
briefly review ECSK cosmology models to make $\Theta^{+}\left(  z\right)  $ estimations for $z>1$.

\subsection{Review of ECSK cosmology}

Let us consider the Copernican ansatz of Eq.~(\ref{Eq_FRW_metric})-(\ref%
{Eq_FRW_Torsion}).
 For a universe with cosmological constant, dark matter and Standard
Model matter, the resulting ECSK field equations are the following
\begin{eqnarray}
3\mathcal{H}^{2} &=&\kappa _{4}c^{2}\rho \,,  \label{pm8} \\
2\mathcal{\dot{H}}-3\mathcal{H}^{2}+2\nu _{+}\mathcal{H} &=&-\kappa
_{4}c^{2}p\,,  \label{pm9}
\end{eqnarray}%
with%
\begin{equation}
\mathcal{H}=H-\nu _{+}\,,
\end{equation}%
and $H=\frac{\dot{a}}{a}$ is the Hubble parameter. The total density and pressure are given by%
\begin{align}
\rho & =\frac{\Lambda }{\kappa _{\mathrm{4}}}+\rho _{\mathrm{SM}}+\rho _{%
\mathrm{DM}}+\frac{3}{c^{2}\kappa _{\mathrm{4}}}\nu _{-}^{2}\,, \\
p& =-\frac{\Lambda }{\kappa _{\mathrm{4}}}+p_{\mathrm{SM}}+p_{\mathrm{DM}}-%
\frac{1}{c^{2}\kappa _{\mathrm{4}}}\nu _{-}^{2}\,.
\end{align}
Assuming dark matter as a source for spin and consequently for torsion,
implies that $\nu _{+}$ and $\nu _{-}$ must be functions of $\rho _{\mathrm{%
DM}}$. However, both functions play a completely different role in Eq.~(\ref%
{pm8})-(\ref{pm9}). It is clear that $\nu _{-}$ acts as an extra
dark matter density and pressure, while $\nu _{+}$ plays a different role in
the dynamics. For this reason, it is natural to assume a \textquotedblleft
barotropic\textquotedblright\ ansatz for $\nu _{-}$ such that%
\begin{equation}
\nu _{-}^{2}=\alpha _{-}^{2}\frac{c^{2}\kappa _{\mathrm{4}}}{3}\rho _{%
\mathrm{DM}}\,,
\end{equation}%
where $\alpha _{-}$ represents a \textquotedblleft
barotropic\textquotedblright\ parameter, providing us with an effective dark
matter density $\rho _{\mathrm{eff}}$ and pressure $p_{\mathrm{eff}}$ given
by%
\begin{eqnarray}
\rho _{\mathrm{eff}} &=&\left( 1+\alpha _{-}^{2}\right) \rho _{\mathrm{DM}}\,,
\\
p_{\mathrm{eff}} &=&\omega _{\mathrm{eff}}\rho _{\mathrm{eff}}\,,
\end{eqnarray}%
with%
\begin{equation}
\omega _{\mathrm{eff}}=\frac{1}{1+\alpha _{-}^{2}}\omega _{\mathrm{DM}}-%
\frac{1}{3}\frac{1}{(1+\frac{1}{\alpha _{-}^{2}})}\,,
\end{equation}%
where have assumed a barotropic relation $p_{\mathrm{DM}}=\omega _{\mathrm{DM%
}}\rho _{\mathrm{DM}}$. We can see that the net effect of $\nu _{-}$ was
only to shift the barotropic constant. In a cold DM case, $\omega _{\mathrm{%
DM}}=0$, we are left with $-\frac{1}{3}<\omega _{\mathrm{eff}}\leq 0$ which
in fact could explain the Hubble parameter tension with values as small as $%
\omega _{\mathrm{eff}}\sim -10^{-2}$, see Ref.~\cite{Izaurieta:2020xpk}.

Equation~(\ref{pm8}, \ref{pm9}) can be rewritten as 
\begin{align}
3\left( H-\nu _{+}\right) ^{2}& =\kappa _{4}c^{2}\rho \,, \\
\dot{\rho}+3H\left( \rho +p\right) -\nu ^{+}\left( \rho +3p\right) & =0\,,
\label{pm10}
\end{align}%
with 
\begin{align}
\rho & =\frac{\Lambda }{\kappa _{\mathrm{4}}}+\rho _{\mathrm{SM}}+\rho _{%
\mathrm{eff}}\,, \\
p& =-\frac{\Lambda }{\kappa _{\mathrm{4}}}+p_{\mathrm{SM}}+p_{\mathrm{eff}}\,.
\end{align}%
In terms of the redshift, Eq.~(\ref{pm10}) becomes%
\begin{equation}
-\left( 1+z\right) \frac{\mathrm{d}\rho }{\mathrm{d}z}+3\left( \rho
+p\right) -\frac{\nu ^{+}}{H}\left( \rho +3p\right) =0\,.
\label{Eq_cosmo_models}
\end{equation}
To solve these equations, we need to know $\nu _{+}$ as a function of the dark
matter density. This is precisely the information that should provide the
field equation~(\ref{deoc2}). However, since we do not have a dark matter
Lagrangian, we have no information a priory on its possible spin tensor and the field
equation~(\ref{deoc2}) becomes useless.

For this reason, when considering ECSK cosmologies, we must propose a
\textquotedblleft reasonable\textquotedblright\ ansatz in order to model the
dependence of $\nu _{+}$ on dark matter density. In the following
sections we briefly review a couple of ansatz that agree with observations,
and we will see how the GW amplitude propagate on them from mergers at $z>1$.

\subsection{Ansatz $\nu _{+}\propto H\rho _{\mathrm{eff}}^{n}$}

Let us consider an ansatz of the form%
\begin{equation}
\nu ^{+}=CH\left( \frac{\rho _{\mathrm{eff}}}{\rho _{\mathrm{c}}}\right)
^{n}\,,  \label{pm12}
\end{equation}%
where $H$ is the Hubble parameter, $\rho _{\mathrm{c}}$ corresponds to the
current critical density%
\begin{equation}
\rho _{\mathrm{c}}=\frac{3H_{0}^{2}}{c^{2}\kappa _{4}}\,,
\end{equation}%
and $C$ is just a proportionality constant.
In general, torsional effects accelerate the expansion of the universe and
may play the role of dark energy. Therefore, let assume a toymodel with $%
\Lambda =0$ and where $\rho _{\mathrm{SM}}$ is negligible when compared to $%
\rho _{\mathrm{eff}}$. In this case and considering the ansatz Eq.~(\ref%
{pm12}), equation (\ref{Eq_cosmo_models}) becomes%
\begin{equation}
-\left( 1+z\right) \frac{\mathrm{d}\rho _{\mathrm{eff}}}{\mathrm{d}z}-\frac{C%
}{\rho _{\mathrm{c}}^{n}}\left( 1+3\omega _{\mathrm{eff}}\right) \rho _{%
\mathrm{eff}}^{n+1}+3\left( 1+\omega _{\mathrm{eff}}\right) \rho _{\mathrm{%
eff}}=0\,.
\end{equation}
The solution of this Bernoulli ODE is (see Ref.~\cite{Pereira:2019yhu,Guimaraes:2020drj})%
\begin{equation}
\rho _{\mathrm{eff}}\left( z\right) =\rho _{\mathrm{c}}\left[ \frac{\frac{3}{%
C}\frac{1+\omega _{\mathrm{eff}}}{1+3\omega _{\mathrm{eff}}}}{\left( \frac{3%
}{C}\frac{1+\omega _{\mathrm{eff}}}{1+3\omega _{\mathrm{eff}}}\frac{1}{%
\Omega _{0}^{n}}-1\right) \left( 1+z\right) ^{-3n\left( 1+\omega _{\mathrm{%
eff}}\right) }-1}\right] ^{\frac{1}{n}}
\end{equation}%
with 
\begin{equation}
\Omega _{0}=\frac{\rho _{\mathrm{eff}0}}{\rho _{\mathrm{c}}}\,.
\end{equation}

Using Eq.~(\ref{pm8}) and~(\ref{pm12}), it is straightforward to find%
\begin{equation}
\frac{H}{H_{0}}=\sqrt{\frac{\rho _{\mathrm{eff}}}{\rho _{\mathrm{c}}}}\frac{1%
}{1-C\left( \frac{\rho _{\mathrm{eff}}}{\rho _{\mathrm{c}}}\right) ^{n}}\,.
\end{equation}%
and for $\frac{\nu ^{+}}{H}$, we have%
\begin{eqnarray}
\frac{\nu ^{+}}{H} &=&C\left( \frac{\rho _{\mathrm{eff}}}{\rho _{\mathrm{c}}}%
\right) ^{n} \\
&=&\frac{3\frac{1+\omega _{\mathrm{eff}}}{1+3\omega _{\mathrm{eff}}}}{\left( 
\frac{3}{C}\frac{1+\omega _{\mathrm{eff}}}{1+3\omega _{\mathrm{eff}}}\frac{1%
}{\Omega _{0}^{n}}-1\right) \left( 1+z\right) ^{-3n\left( 1+\omega _{\mathrm{%
eff}}\right) }-1}\,. \label{pmm1}
\end{eqnarray}
The Ref.~\cite{Pereira:2019yhu} shows that the observations allow for the following
ranges of parameters for this model:%
\begin{table}[H]
\begin{center}
\caption {\textbf{Observational Parameters}}
\label{tab:data} 
\begin{tabular}
[c]{cccc}\hline\hline
$\Omega_{0}$ & $C$ & $n$ & $\underset
{(\mathrm{\frac{km}{s}Mpc})}{H_{0}}$\\ \hline
$0.31_{-0.12}^{+0.11}$ & $0.28_{-0.24}^{+0.28}$ & $-0.47_{-0.36}^{+0.26}$ &
$68.8_{-3.1}^{-3.0}$\\ \hline
\end{tabular}
\end{center}
\end{table}
Inserting Eq.~(\ref{pmm1}) into  Eq.~(\ref{Eq_Theta+Z}),  with the approximation $\omega _{\mathrm{eff}}\approx 0$ (cold dark matter), we find%
\begin{equation}
\Theta ^{+}\left( z\right) =-\int_{z}^{0}\mathrm{d}\tilde{z}\,\frac{1}{1+%
\tilde{z}}\frac{3}{\left( \frac{3}{C}\sqrt{\Omega _{0}}-1\right) \left( 1+%
\tilde{z}\right) ^{-3n}+1}\,.
\end{equation}
From the allowed observational parameters Table \ref{tab:data}, we can make an estimation for $\Theta ^{+}\left( z\right)$. In fact, this function grow too slowly to compensate for the smallness of $\frac{1}{2}%
\left\vert P_{0}\right\vert ^{2}$ in Eq.~(\ref{Eq_eA_Pchico}). To see this, it suffices
to make a plot and to compare it with
the estimates of Eq.~(\ref{Eq_Theta_MIN}).

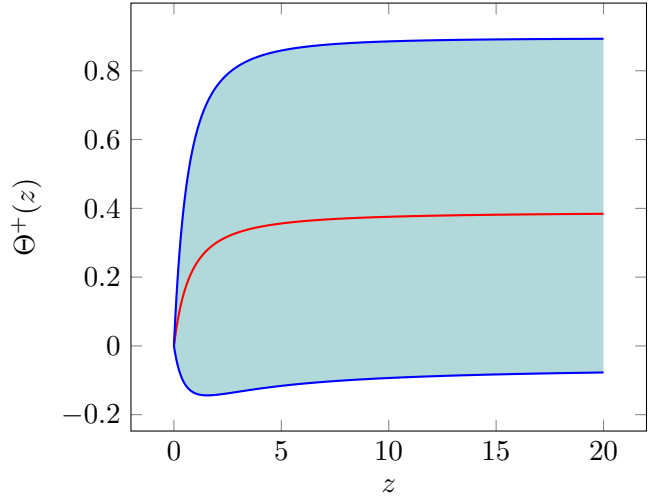
\begin{figure}[H]
	\begin{center}
		\begin{tikzpicture}
		\begin{axis}[
		axis lines = box,
		xlabel = \(z\),
		ylabel = {\(\Theta^{+}(z)\)},
		]
		\addplot[color=red,thick, smooth, domain=0:20,samples=400]{(1/0.47)*ln(5.97/(4.97 + (1 + x)^(-3*0.47)))};
		\addplot[name path=Up, color=blue, thick,smooth,domain=0:20,samples=400]
		{1/0.47*ln(5.97/(4.97 + (1 + x)^(-3*0.47))) + 7.02*abs(1/0.47*((1 + x)^(-3*0.47) - 1)/((4.97 + (1 + x)^(-3*0.47))*5.97)) + 0.26*abs(-3*1/0.47*(1 + x)^(-3*0.47)*ln(1 + x)/(4.97 + (1 + x)^(-3*0.47)))};
		\addplot[name path=Low, color=blue, thick, smooth, domain=0:20,samples=400]
		{1/0.47*ln(5.97/(4.97 + (1 + x)^(-3*0.47))) - 6.27*abs(1/0.47*((1 + x)^(-3*0.47) - 1)/((4.97 + (1 + x)^(-3*0.47))*5.97)) - 0.36*abs(-3*1/0.47*(1 + x)^(-3*0.47)*ln(1 + x)/(4.97 + (1 + x)^(-3*0.47)))};
		\addplot [teal!30] fill between[of=Up and Low];
		\end{axis}
		\end{tikzpicture}
		\caption{$\Theta ^{+}\left( z\right)$ as a function of redshift $z$. Blue lines contour the upper and lower deviation from the mean values showed in red. }
	\end{center}
\end{figure}

\subsection{Ansatz $\protect\nu _{+}\propto H$}

The steady-state torsion ansatz, i.e.,
\begin{equation}
\nu _{+}=-\alpha H\,,
\end{equation}
with $\alpha $ constant, is very popular among the ECSK cosmology models
Ref.~\cite{Pereira:2022cmu}. The main reason for this is that Eq.~(\ref{Eq_cosmo_models})
becomes easily integrable under this ansatz.
However, it has a severe physical drawback: from the field equation~(\ref%
{deoc2}) we expect that $\nu _{+}$ should depend on some dark matter feature,
e.g., its density, and not on a geometrical feature as $H$.
Regardless of whether or not we should consider $\nu _{+}=-\alpha H$ as a
realistic astrophysical model, let us assess its capacity to give rise to
some observable effect on GW amplitude propagation.
In this case, we have that Eq.~(\ref{Eq_Theta+Z}) becomes
\begin{eqnarray}
\Theta ^{+}\left( z\right)  &=&\alpha \int_{z}^{0}\mathrm{d}\tilde{z}\,\frac{%
1}{1+\tilde{z}}\,, \\
&=&-\alpha \ln \left( 1+z\right) \,.
\end{eqnarray}
According to Ref.~\cite{Pereira:2022cmu}, the observations constrain this model
allowing for a small positive value for $\alpha $, 
\begin{equation}
\alpha =0.086_{-0.095}^{+0.094}\,.
\end{equation}
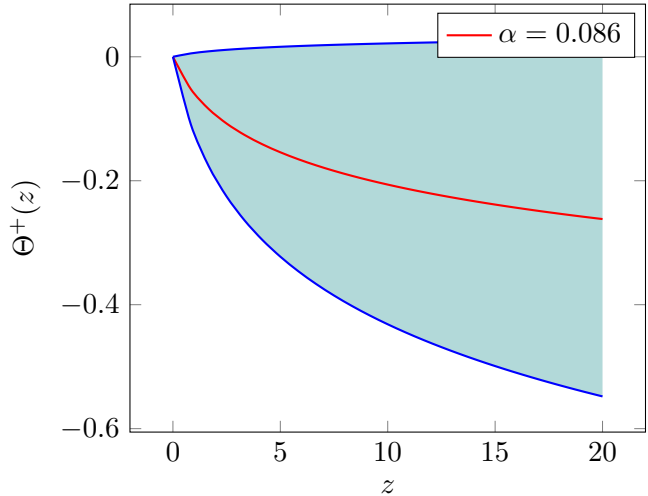
\begin{figure}[H]
\begin{center}
\begin{tikzpicture}
\begin{axis}[
    axis lines = box,
    xlabel = \(z\),
    ylabel = {\(\Theta^{+}(z)\)},
]
    \addplot[color=red,thick, smooth, domain=0:20]{-(0.086)*ln(1+x)};
    \addlegendentry{$\alpha=0.086$}
    \addplot[name path=f1, color=blue,thick,smooth, domain=0:20]
    {-(0.086+0.094)*ln(1+x)};
    \addplot[name path=f2, color=blue,thick,smooth,domain=0:20]
    {-(0.086-0.095)*ln(1+x)};
    \addplot [teal!30] fill between[of=f1 and f2];
\end{axis}
\end{tikzpicture}
\caption{$\Theta ^{+}\left( z\right)$ as a function of redshift $z$. }
\end{center}
\end{figure}
This negative (and slow-evolving) $\Theta ^{+}\left( z\right) $ stands no
chance to compensate for the smallness of $\frac{1}{2}\left\vert
P_{0}\right\vert ^{2}$ in Eq.~(\ref{Eq_eA_Pchico}), this $\Theta ^{+}\left(
z\right) $ is well below the minimal value of Eq.~(\ref{Eq_Theta_MIN}).


\section{Conclusions and Outlook}\label{SecV}
After performing a general analysis of gravitational waves on Riemann-Cartan geometries
and their associated features (reviewed in Sec.~\ref{SecII} and Ref.~\cite{Barrientos:2019msu,Valdivia:2017sat,Barrientos:2019awg}), it becomes clear that
a torsional background would generically produce an anomalous propagation of a
GW amplitude and polarization, as given by Eq.~(\ref{Eq_Non_Consv_J})-(\ref{Eq_Polarization_Propag}). In particular, in the present article we have focused
our attention on mergers' GW amplitude propagation in the context of an ECSK theory,
where a nonvanishing dark matter spin tensor could behave as a possible
torsion source on cosmological scales.

Given the solid observational success of standard GR, we consider a
\textquotedblleft weak-torsion scenario.\textquotedblright\ This means that
the background torsion is set to be weak enough to make the mergers GW emission
process indistinguishable (at least at leading and subleading orders) from
the one predicted by standard GR. 

This leads, in particular, to a hierarchy of polarization modes, where at the
moment of emission, polarization modes $\left( b\right) $, $\left(
l\right) $, $\left( x\right) $, and $\left( y\right) $ are at least $%
\epsilon $ times weaker than the modes $\left( +\right) $ and $\left( \times
\right) $ (see Eq.~(\ref{Eq_hierarch_1})-(\ref{Eq_hierarch_epsilon2})).

However, the anomalous propagation of polarization in Eq.~(\ref{Eq_Polarization_Propag}) implies (at least in principle) that torsion could slowly
amplify these weaker polarization modes along the geodesic. In the same way, and
also in principle, background torsion could lead to an anomalous dampening (or
amplification) of the amplitude, Eq.~(\ref{Eq_Non_Consv_J})-(\ref{Eq_phi vs
phi}), after going across a long cosmological distance.

Since the merger masses and angular momentum correlate to the GW frequency
evolution, an anomalous amplitude propagation could lead to a wrong
luminosity distance assessment $\mathring{D}_{\mathrm{L}}$ (see Eq.~(\ref{Eq_D/D})), potentially disabling the use of mergers as standard candles.

The main result in this paper is the conclusion, following from our analysis, that these worries are actually unfounded, so that mergers remain to be reliable standard candles, both with or without torsion (at least for
ECSK theories). In Secs.~\ref{SecIII} and~\ref{SecIV}  the strength of this
anomalous amplitude propagation has been calculated for a generic ECSK theory. The conclusion is that the effect of torsion under these conditions is so tiny that it remains always below detection thresholds, even thinking in near-future interferometers, such as LISA. In particular, for the ECSK models
considered in Sec.~\ref{SecIV}, to detect a possible anomalous amplitude propagation
due to torsion one would need to measure mergers luminosity distances at $z=1
$ with a precision of
\begin{equation}
\frac{\delta \mathring{D}_{\mathrm{L}}}{\mathring{D}_{\mathrm{L}}}\leq
10^{-41}.
\end{equation}%
To detect such an effect may still be possible, but it is as of now
beyond technological capabilities.

Elaborating further on this point, Eq.~(\ref{Eq_eA_Pchico}) shows why this is the case. Even if the anomalous
amplitude effect accumulates over a long cosmological distance in the integral 
$\Theta ^{+}$, the smallness of the GW trace modes $\left( b\right) $ and $%
\left( l\right) $ at the moment of emission renders the whole integrated effect still
negligible. In turn, the smallness of $\frac{1}{2}\left\vert
P_{0}\right\vert ^{2}$ is a direct consequence of the weak-torsion scenario,
where torsion should be kept weak enough so that the GW emission process still happens as in the well checked  standard GR.

This point opens a critical issue for further research. First, it is crucial
to notice that the weak-torsion hypothesis is reasonable, but  for the late
universe only. In contrast, at the high spin densities of the very early universe,
torsion could actually have been relevant \cite{Poplawski:2011jz,Poplawski:2012qy,Cubero:2019lxw, Poplawski:2020hrp, Poplawski:2020kzc}, and the GW emission process
could have significantly departed, at that epoch, from the one predicted by GR \cite{Battista:2021rlh}.
In this case, it is not clear yet whether $\frac{1}{2}\left\vert
P_{0}\right\vert ^{2}$ in Eq.~(\ref{Eq_eA_Pchico}) should still be
considered negligible. Therefore, anomalous propagation of amplitude and
polarization, Eq.~(\ref{Eq_Non_Consv_J})-(\ref{Eq_Polarization_Propag}),
produced by torsion could, in principle, leave a detectable fingerprint in
the cosmic gravitational wave background. This promising topic will be covered
elsewhere.

\section{Acknowledgements}

We are grateful to
Sergei Odintsov,
Crist\'{o}bal Corral,
Patricio Salgado,
Sebasti\'{a}n Salgado,
and Jorge Zanelli,
for many enlightening conversations.
This work has been partially supported by MICINN (Spain), project PID2019-104397GB-I00 of the Spanish State Research Agency program AEI/10.13039/501100011033, by the Catalan Government, AGAUR project 2017-SGR-247, and by the program Unidad de Excelencia María de Maeztu CEX2020-001058-M.

FI acknowledges financial support from the Chilean government through Fondecyt grants 1150719, 1180681 and 1211219. FI is thankful of the emotional support of the Netherlands Bach Society. They made freely available superb quality recordings of the music of Bach, and without them, this work would have been impossible.
OV acknowledges to ICE-CSIC for the hospitality, VRIIP-UNAP for financial support through Project VRIIP0258-18, Becas Chile ANID project 74200062 and Fondecyt 11200742.

\appendix
\section{Anomalous propagation of amplitude and polarization}\label{appA}


Since
\begin{equation}
\mathring{\nabla}_{\mu}J^{\mu}=\nabla_{\mu}J^{\mu}-T^{\mu}{}_{\mu\lambda}J^{\lambda}\,. \label{dtor}%
\end{equation}
using Eq.~(\ref{Eq_Def_Ampl_Pol}), Eq.~(\ref{Eq_Pol_Norm}) in Eq.~(\ref{Ec_Origen_Amplitud_Polarizacion}), with $J^{\mu}=k^{\mu}\varphi^2$, the ray current conservation breaks according to
\begin{equation}
\mathring{\nabla}_{\lambda}J^{\lambda}=\left(  \Pi^{\mu\nu}-g^{\mu\nu}\right)T_{\mu\nu\lambda}J^{\lambda}\,,%
\end{equation}
where $\Pi^{\mu\nu}$ is given in Eq.~(\ref{Eq_Pi_mn}).

Recall Eq.~(\ref{Ec_Origen_Amplitud_Polarizacion}) written in the form
\begin{equation}
k^{\lambda}\nabla_{\lambda}\mathbb{H}_{\mu\nu}+\frac{1}{2}\left(\nabla_{\lambda}k^{\lambda}\mathbb{H}_{\mu\nu}+k^{\lambda}M_{\lambda\mu\nu}^{+}\right)=0
\end{equation}
Inserting Eq.~(\ref{Eq_Def_Ampl_Pol}) and $J^{a}=\varphi^{2}k^{a}$ one finds
\begin{equation}
k^{\lambda}\nabla_{\lambda}P_{\mu\nu}+\frac{1}{2\varphi^{2}}\nabla_{\lambda}J^{\lambda}P_{\mu\nu}+\frac{1}{2\varphi}k^{\lambda}M_{\lambda\mu\nu}^{+}=0\,.\label{corcho}
\end{equation}
Using the fact that $\nabla_{\lambda}J^{\lambda}=\Pi^{\mu\nu}T_{\mu\nu\lambda}J^{\lambda}$ direct calculation shows
\begin{equation}
\nabla_{\lambda}P_{\mu\nu}=\mathring{\nabla}_{\lambda}P_{\mu\nu}+\frac{1}{2}\left(T_{\rho\nu\lambda}-T_{\nu\rho\lambda}+T_{\lambda\nu\rho}\right)  P^{\rho}{}_{\mu}+\frac{1}{2}\left(  T_{\rho\mu\lambda}-T_{\mu\rho\lambda}+T_{\lambda\mu\rho}\right)  P^{\rho}{}^{\nu}
\end{equation}
Using this result we can finally solve for $k^{\lambda}\mathring{\nabla}_{\lambda}P_{\mu\nu}$ as
\begin{align}
k^{\lambda}\mathring{\nabla}_{\lambda}P_{\mu\nu}  & =-\frac{1}{2}k^{\lambda
}\left[  \Pi^{\rho\sigma}T_{\rho\sigma\lambda}P_{\mu\nu}+\left(
T_{\sigma\lambda\nu}-\frac{1}{2}\left[  T_{\lambda\sigma\nu}+T_{\nu
	\sigma\lambda}\right]  \right)  P_{~\mu}^{\sigma}\right.  +\nonumber\\
& \left.  \left(  T_{\sigma\lambda\mu}-\frac{1}{2}\left[  T_{\lambda\sigma\mu
}+T_{\mu\sigma\lambda}\right]  \right)  P_{~\nu}^{\sigma}+\frac{1}{2}g_{\mu
	\nu}T_{\rho\sigma\lambda}P^{\rho\sigma}-\frac{1}{4}\left[  g_{\mu\nu
}T_{~\sigma\lambda}^{\sigma}-\left(  T_{\mu\nu\lambda}+T_{\nu\mu\lambda
}\right)  \right]  P\right]%
\end{align}
\bibliography{GB2019}
\bibliographystyle{utphys}

\end{document}